%% file: main.tex
\documentclass[journal, hidelinks]{IEEEtran}
\usepackage{cite}
\usepackage{amsmath,amssymb,amsfonts}
\usepackage{algorithmic}
\usepackage{tabularx}
\usepackage{comment}
\usepackage{booktabs}
\usepackage{threeparttable}
\usepackage{multirow}
\usepackage{siunitx}
\usepackage{relsize}
\usepackage{textcomp}
\usepackage{xcolor}
\usepackage{amsmath, empheq}
\usepackage[most]{tcolorbox}
\usepackage{graphicx}
\usepackage{booktabs}
\usepackage{arydshln}
\usepackage{threeparttable}
\usepackage{mathrsfs}
\usepackage{multirow}
\usepackage{hyperref}
\usepackage{graphicx}
\allowdisplaybreaks

\usepackage{eucal}

\newtcbox{\mymath}[1][]{%
    nobeforeafter, 
    math upper, 
    tcbox raise base,
    enhanced, 
    colback=blue!5!white, 
    colframe=blue!20!white, 
    boxrule=0.7pt,
    height=4.5em, 
    width=6em,  
    left=-0.01em,  
    right=-0.01em, 
    #1
}

\usepackage{amsfonts,amsmath,amssymb,amsthm}
\usepackage[super]{nth}

\usepackage{rotating} 

\usepackage{algorithmic}
\usepackage[ruled,linesnumbered,noresetcount,vlined]{algorithm2e}

\ifCLASSOPTIONcompsoc
 \usepackage[caption=false,font=normalsize,labelfont=sf,textfont=sf]{subfig}
\else
 \usepackage[caption=false,font=footnotesize]{subfig}
\fi

\SetKwInput{KwInput}{Input}                
\SetKwInput{KwOutput}{Output}

\usepackage{float}
\floatstyle{ruled}
\newfloat{model}{thp}{lop}
\floatname{model}{Model}

\usepackage{enumitem}
\newlist{abbrv}{itemize}{1}
\setlist[abbrv,1]{label=,labelwidth=0.5in,align=parleft,itemsep=0.1\baselineskip,leftmargin=!}

\usepackage{booktabs}
\usepackage{csvsimple}
\usepackage{siunitx}

\usepackage{comment}
\newif\ifarxiv
\arxivtrue
\arxivfalse

\usepackage{orcidlink}

\renewcommand{\baselinestretch}{1}

\DeclareMathOperator*{\minimize}{minimize}

\begin{document}

\title{Optimizing Parameters of the LinDistFlow Power Flow Approximation for Distribution Systems}

\author{\IEEEauthorblockN{
Babak Taheri \orcidlink{0000-0002-7870-3465},
Rahul K. Gupta \orcidlink{0000-0002-9905-6092}, and
Daniel K. Molzahn \orcidlink{0000-0003-0583-5376}
\thanks{\noindent Electrical and Computer Engineering, Georgia Institute of Technology. \{taheri, rahul.gupta, molzahn\}@gatech.edu. Support from NSF \#2145564.}
}
}

\maketitle

\begin{abstract}
The \mbox{DistFlow} model accurately represents power flows in distribution systems, but the model's nonlinearities result in computational challenges for many applications. Accordingly, a linear approximation known as \mbox{LinDistFlow} (and its three-phase extension \mbox{LinDist3Flow}) is commonly employed. This paper introduces a parameter optimization algorithm for enhancing the accuracy of this approximation for both balanced single-phase equivalent and unbalanced three-phase distribution network models, with the goal of aligning the outputs more closely with those from the nonlinear \mbox{DistFlow} model. Using sensitivity information, our algorithm optimizes the \mbox{LinDistFlow} approximation's coefficient and bias parameters to minimize discrepancies in predictions of voltage magnitudes relative to the nonlinear \mbox{DistFlow} model. 
The parameter optimization algorithm employs the Truncated Newton Conjugate-Gradient (TNC) method to fine-tune coefficients and bias parameters during an offline training phase to improve the \mbox{LinDistFlow} approximation's accuracy.
Numerical results underscore the algorithm's efficacy, showcasing accuracy improvements in $L_{1}$-norm and $L_{\infty}$-norm losses of up to $92\%$ and $88\%$, respectively, relative to the traditional \mbox{LinDistFlow} model.
We also assess how the optimized parameters perform under changes in the network topology and demonstrate the optimized \mbox{LinDistFlow} approximation's efficacy in a hosting capacity optimization problem.

\end{abstract}


\input{text/introduction2}

\input{text/LinDistFlow}

\input{text/proposed_algorithm}

\input{text/results}

\section{Conclusion}
\label{sec:conclusion}

The \mbox{LinDistFlow} approximation is often used to improve the computational tractability of optimization problems in distribution systems. This paper introduces a new algorithm that enhances the accuracy of the \mbox{LinDistFlow} approximation for single- and three-phase network models. Inspired by machine learning methods, the algorithm's offline phase optimizes the approximation's coefficients and biases using analytically derived sensitivities within the TNC optimization method. These optimized parameters then provide increased accuracy when used in various applications. Numerical tests demonstrate the algorithm's effectiveness, showing better alignment with nonlinear DistFlow solutions compared to the traditional \mbox{LinDistFlow} and other recent alternatives. Application to a hosting capacity problem further highlights its advantages.

Since our optimized formulation has the same linear form as the traditional \mbox{LinDistFlow} approximation and performs well for varying topologies, the accuracy advantages of the proposed approach can be directly exploited in a wide range of other applications such as those in~\cite{baran1989optimal1,baran1989optimal2,baran1989network,baker2017network, arnold2016optimal,kekatos2016voltage,robbins2016optimal,Mieth2018,gupta2024improving,bose2023lindistflow}. Our future work aims to explore such applications.  
\textcolor{black}{Building on prior literature such as~\cite{wu2017oltc,bazrafshan2018pscc,bazrafshan2019}, our future work will also focus on extending the OLDF framework to more effectively optimize setpoints for active control devices, such as voltage regulators and switchable capacitor banks. While the current OLDF enhances the modeling of the passive network (with capacitor bank actions represented by their reactive power injections, $q_n$, and fixed regulator effects implicitly learned during training), its optimal integration with dynamically operating controls warrants dedicated investigation. Key research directions include tailoring OLDF training strategies to capture the diverse system states induced by these controls and developing methods for adapting OLDF parameters based on their operational status, potentially drawing insights from our topology adaptation analysis. Such advancements aim to further enhance OLDF's practical utility in actively controlled distribution systems. 
}

\bibliographystyle{IEEEtran}
\bibliography{blib}

\input{text/appendix}

\begin{IEEEbiography}[{\includegraphics[width=1in, height=1.25in, clip, keepaspectratio]{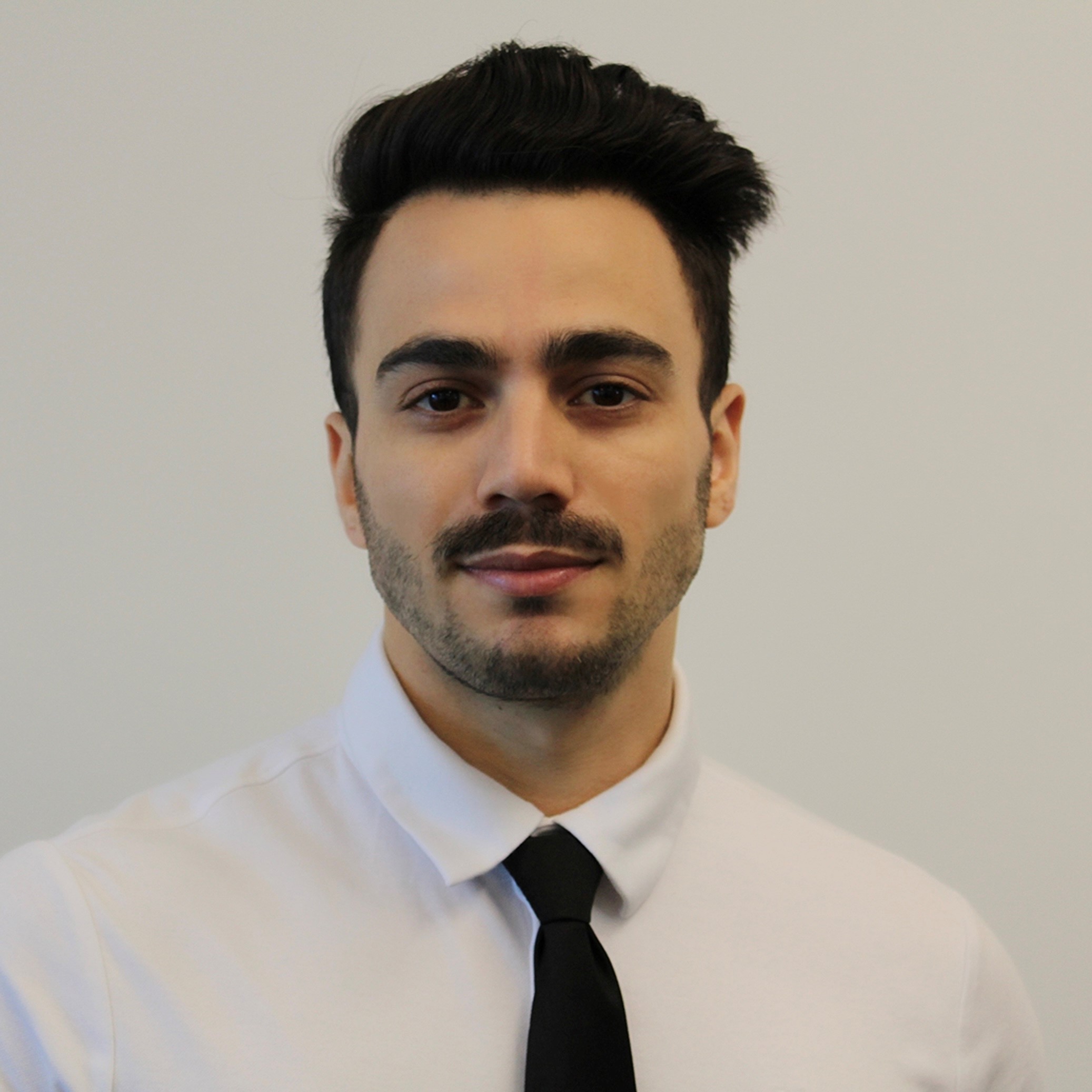}}]{Babak Taheri} received the Ph.D. and M.Sc. degrees in electrical and computer engineering from the Georgia Institute of Technology, Atlanta, GA, USA in 2024. He also received the M.Sc. degree in electrical engineering from Sharif University of Technology in 2019 and the B.Sc. degree from the University of Tabriz in 2017. He is currently a Research Scientist at Hitachi Energy Research. His research interests include power systems, optimization, and machine learning.
\end{IEEEbiography}

\vspace{-7pt}

\begin{IEEEbiography}[{\includegraphics[width=1in, height=1.25in, clip, keepaspectratio]{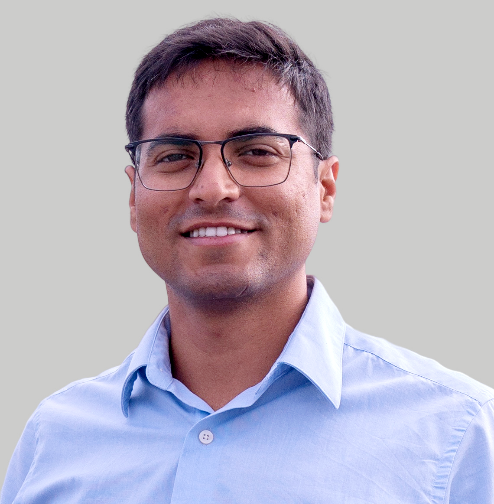}}]{Rahul K. Gupta} 
(Member, IEEE) received
the B.Tech. degree in electrical engineering
from the National Institute of Technology
(NIT), Rourkela, India, in 2014, and the M.Sc.
and Ph.D. degrees in electrical engineering
from the Swiss Federal Institute of Technology
(EPFL), Lausanne, Switzerland, in 2018
and 2023, respectively.
From October 2023 to December 2024,
he was a Postdoctoral Scholar with the School of Electrical and Computer Engineering, Georgia Institute of Technology, Atlanta, GA, USA, supported by a grant from the Swiss National Science Foundation (SNSF). Since 2025, he has been an Assistant Professor with the School of Electrical Engineering and Computer Science
(EECS), Washington State University (WSU), Pullman, WA, USA. 
He was the recipient of the ABB Research Award 2025 and EPFL Ph.D. Thesis Distinction in Electrical Engineering for his doctoral research.

\end{IEEEbiography}

\vspace{-7pt}

\begin{IEEEbiography}[{\includegraphics[width=1in, height=1.25in, clip, keepaspectratio]{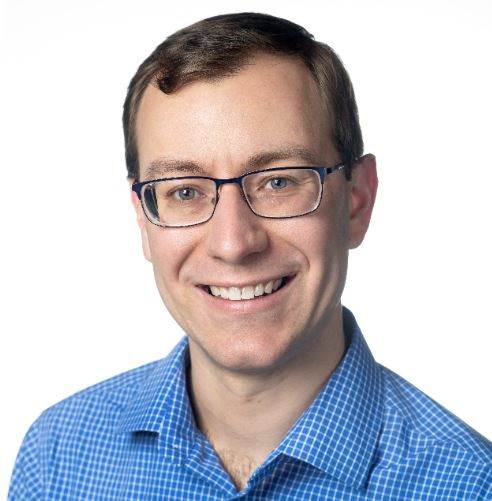}}]{Daniel K. Molzahn} (Senior Member, IEEE) received
the B.S., M.S., and Ph.D. degrees in electrical engineering and the master’s of Public Affairs degree from the University of Wisconsin--Madison, Madison, WI, USA. He is currently an Associate Professor with the School of Electrical and Computer Engineering, Georgia Institute of Technology, Atlanta, GA, USA, and also holds an appointment as a Computational
Engineer in the Energy Systems Division at Argonne National Laboratory. He was a Dow Postdoctoral Fellow in Sustainability at the University of Michigan, Ann Arbor, MI, USA, and a National Science Foundation Graduate Research Fellow at the University of Wisconsin--Madison. He was the recipient of the IEEE Power and Energy Society’s Outstanding Young Engineer Award in 2021, the NSF CAREER Award in 2022, and Georgia Tech’s Class of 1940 W. Roane Beard Outstanding Teacher Award in 2024.
\end{IEEEbiography}

\end{document}

%% file: text/introduction2.tex
\section{Introduction}
\label{sec:Introduction}
Power flow models relating power injections, line flows, and voltages are central to power system design and operation~\cite{stott1974review}. 
\textcolor{black}{While all mathematical models of physical systems are inherently approximations, the AC power flow equations are the established, high-fidelity benchmark used to accurately model these relationships.}
Incorporating the AC power flow equations into optimization problems introduces significant computational challenges due to these equations' nonlinearity, even for the radial networks of typical distribution systems~\cite{pascalNPhard}. These challenges are particularly relevant to problems that are run online to inform real-time decisions~\cite{overbye2004comparison,misra2017irep}, consider uncertainties~\cite{pscc2022_survey}, and model discrete decisions
~\cite{barrows2014,castillo2016,austgen2023comparisons,rhodes2023}. To address these challenges, engineers frequently use power flow approximations that trade accuracy for tractability~\cite{molzahn2019}. 

Distribution systems are often modeled using an AC power flow formulation known as ``DistFlow''~\cite{baran1989optimal1, baran1989optimal2, baran1989network}. A linearization of the \mbox{DistFlow} model known as \mbox{``LinDistFlow''}~\cite{baran1989optimal1,baran1989optimal2,baran1989network}, or one of its \textcolor{black}{variants~\cite{gan2014convex, kekatos2016voltage, robbins2016optimal,yang2017linearized,schweitzer2020lossy,markovic2022parameterized, Madani2025SOCP_DistFlow},} is commonly employed to make distribution system optimization problems tractable. An extension known as \mbox{LinDist3Flow} is often used for unbalanced three-phase distribution network models~\cite{gan2014convex}.
The traditional \mbox{LinDistFlow} approximation linearizes the \mbox{DistFlow} equations by assuming that the active and reactive line losses are much smaller than the active and reactive line flows. \mbox{LinDistFlow} has been used to site and size capacitors~\cite{baran1989optimal1,baran1989optimal2}, optimize the network topology~\cite{baran1989network}, compute inverter setpoints and perform Volt/VAr control~\cite{baker2017network, arnold2016optimal,kekatos2016voltage,robbins2016optimal}, formulate stochastic problems~\cite{Mieth2018}, ensure fairness in solar photovoltaic curtailments~\cite{gupta2024improving}, and calculate electricity prices~\cite{bose2023lindistflow}, among many other applications.
Despite their extensive applications, the accuracy of such models can vary, particularly outside near-nominal operating regions, as indicated in~\cite{robbins2016optimal, chen2018robust}. \textcolor{black}{Recent work continues to examine the inherent errors in the LinDistFlow model and their impact on applications such as flexibility aggregation, proposing compensation methods to refine its accuracy~\cite{Dai2025Flexibility}}. \textcolor{black}{This variability in accuracy poses a significant challenge, as reliance on potentially inaccurate models can lead to suboptimal operational decisions, misjudgment of system limits (such as hosting capacity), and inefficient investment in distribution system infrastructure. Therefore, enhancing the fidelity of these linear approximations without compromising their computational tractability is of considerable practical importance for the reliable and economic operation and planning of modern distribution networks.}

Building on ideas from recently developed ``adaptive'' power flow approximations~\cite{misra2018optimal,hohmann2018,muhlpfordt2019optimal,liu2019data,liu2022data,chen2023naps,buason2022sample,taheri2023optimizing, jia2023overview, jia2023tutorial1, jia2023tutorial2, taheri2024improving,rosemberg2025differentiable, chen2025optimally, taheri2024acinformed}, 
\textcolor{black}{and related advancements focusing on areas like second-order sensitivity insights~\cite{buason2024adaptive} and sample-based piecewise linear approaches~\cite{cho2024datadriven,buason2025sample},}
\textcolor{black}{this paper introduces a parameter optimization algorithm for the \mbox{LinDistFlow} model (and its three-phase extension, \mbox{LinDist3Flow}) to significantly enhance its accuracy. Our method optimizes key coefficient parameters, specifically the elements of the diagonal matrices $D_r$ and $D_x$ which represent effective line resistances and reactances, and newly introduced bias parameters. These bias parameters are vector offsets $\boldsymbol{\rho}$ for active power injections ($p$), $\boldsymbol{\varrho}$ for reactive power injections ($q$), and $\boldsymbol{\gamma}$ for squared voltage magnitudes ($v$). These parameters, which form the basis of our Optimized \mbox{LinDistFlow} (OLDF) formulation detailed further in Section~\ref{sec:proposed algorithm}, are tuned via an offline training process. This allows the OLDF model to better emulate the nonlinear DistFlow model, thereby more accurately capturing the net impact of complex physical effects, such as those related to power losses, which are simplified in traditional \mbox{LinDistFlow} formulations.} 

Adaptive power flow approximations like the Optimized \mbox{LinDistFlow} in this paper are linearizations that are tailored to a specific system and operating range of interest. This contrasts with traditional power flow approximations that are often derived using general assumptions about broad classes of systems or are based on a particular nominal operating point. Adaptive power flow approximations invest computing time up front to calculate linearization coefficients in order to achieve increased accuracy when the approximations are deployed in an optimization problem. Thus, adaptive approximations are well suited for settings with both offline and online aspects (e.g., using a day-ahead forecast to compute linearization coefficients that are used for online computations in real-time applications~\cite{overbye2004comparison,misra2017irep}) as well as settings where a nonlinear AC power flow model would lead to an intractable formulation~\cite{pscc2022_survey,barrows2014,castillo2016,austgen2023comparisons,rhodes2023}.

Our prior work in~\cite{taheri2023optimizing} proposes an adaptive DC power flow approximation that optimizes parameters for a particular system over a specified operating range of interest. Other adaptive power flow approximations minimize worst-case error~\cite{misra2018optimal} and expected error~\cite{muhlpfordt2019optimal} or leverage sample-based regression approaches~\cite{liu2019data,liu2022data,chen2023naps} including techniques for constructing overestimating and underestimating approximations~\cite{buason2022sample}. \textcolor{black}{Recent advancements in this area explore the use of second-order sensitivities to both analyze and improve these approximations, guiding the development of, for instance, rational approximations and conservative piecewise linear functions tailored to directions of high curvature~\cite{buason2024adaptive}, as well as sample-based piecewise linear models~\cite{buason2025sample};} 
see~\cite{jia2023overview}
for a recent survey.

\textcolor{black}{Recently, there has been growing interest in developing advanced parameter optimization techniques for various linear power flow models. While our work focuses on the \mbox{LinDistFlow} approximation for distribution systems, several noteworthy approaches have also been proposed for transmission systems, often targeting the DC power flow model and its variants (e.g., \cite{cho2024datadriven,taheri2024improving, taheri2024acinformed,  rosemberg2025differentiable, chen2025optimally, taheri2023optimizing}). These methodologies for transmission systems address different objectives and parameterization scopes, conditioned by the distinct topological and electrical characteristics of high-voltage networks. For instance, recent work by the authors has explored parameter optimization for improving DC Optimal Power Flow (DC-OPF) formulations~\cite{taheri2024improving} and for AC-informed DC optimal transmission switching problems~\cite{taheri2024acinformed} in such transmission system contexts. 
\textcolor{black}{The continued application of \mbox{LinDistFlow} in distribution system analyses, such as in data-driven distributionally robust optimal power flow~\cite{Li2025WassersteinOPF}, underscores the ongoing need for enhancing its accuracy.}}

Contrasting with prior adaptive power flow approximations~\cite{misra2018optimal,muhlpfordt2019optimal,liu2019data,liu2022data,chen2023naps,buason2022sample,jia2023overview},
this paper develops a parameter optimization algorithm that maintains the structure of the \mbox{LinDistFlow} approximation (and the unbalanced three-phase extension \mbox{LinDist3Flow}) as dictated by the network topology. Thus, the resulting parameter-optimized \mbox{LinDistFlow} approximation has the key advantages of being directly deployable in the many existing distribution system applications that rely on \mbox{LinDistFlow} (e.g.,~\cite{baran1989optimal1,baran1989optimal2,baran1989network,baker2017network, arnold2016optimal,kekatos2016voltage,robbins2016optimal,Mieth2018,gupta2024improving,bose2023lindistflow}) and enabling straightforward modeling of topology changes.

While not using traditional machine learning models like neural networks, our parameter optimization algorithm draws inspiration from methods for training machine learning models. We define a loss function that compares the AC power flow solutions to the approximation's outputs over a set of sampled power injections. With analytically calculated parameter sensitivities, an offline training phase computes the parameter values that minimize this loss function using a Truncated Newton Conjugate-Gradient (TNC) method. Using linear programming or mixed-integer linear programming solvers, the optimized parameters are then used in online calculations for real-time settings or in problems for which a nonlinear AC power flow model would lead to intractability. The proposed approach is conceptually similar to our previous work for optimizing the parameters of the DC power flow approximation~\cite{taheri2023optimizing}, but is applicable to unbalanced three-phase distribution systems. 

Numerical comparisons demonstrate substantial accuracy advantages of our proposed algorithm compared to the traditional \mbox{LinDistFlow}/\mbox{LinDist3Flow} approximation as well as several recent \mbox{LinDistFlow} variants that also tune parameter values~\cite{yang2017linearized,schweitzer2020lossy,markovic2022parameterized}. We also study how the approximation accuracy is affected by changes in network topology. Finally, to illustrate the benefits of our algorithm, we apply the optimized \mbox{LinDistFlow} parameters in a hosting capacity analysis.

To summarize, the key contributions of this paper are:
\textcolor{black}{
\begin{itemize}
\item \textbf{A systematic framework for optimizing \mbox{LinDistFlow} and \mbox{LinDist3Flow}} to significantly enhance accuracy to better match nonlinear DistFlow by tuning coefficient and bias parameters using sensitivity analysis and the TNC method.
\item \textbf{A training methodology demonstrating enhanced model robustness} by optimizing \mbox{LinDistFlow} parameters over diverse operating conditions, outperforming traditional fixed-assumption or single-point methods.
\item \textbf{Comprehensive evidence regarding the superior accuracy of the optimized \mbox{LinDistFlow} } (up to $92\%$ $L_1$-norm and $88\%$ $L_\infty$-norm error reductions) over traditional and recent linear variants across diverse test systems, loading conditions, and for both single and three-phase models.
\item \textbf{Insights into the optimized model's adaptability}, showcasing its ability to maintain enhanced accuracy across varying network topologies.
\item \textbf{Demonstration of the \textcolor{black}{proposed OLDF's} efficacy in practical applications}, yielding more reliable and feasible solutions than traditional LDF in complex engineering tasks like hosting capacity analysis.
\end{itemize}
}

The remainder of this paper is organized as follows: Section~\ref{sec:PF} reviews the \mbox{DistFlow} model and \mbox{LinDistFlow} approximation. Section~\ref{sec:proposed algorithm} introduces our proposed algorithm. Section~\ref{sec:Multi-Phase Grids} extends the algorithm to three-phase network models. Section~\ref{sec:Numerical experiments} presents numerical experiments that evaluate the performance of our algorithm. Section~\ref{sec:conclusion} concludes the paper.

%% file: text/LinDistFlow.tex
\section{Power Flow Modeling}
\label{sec:PF}

This section introduces the \mbox{DistFlow} formulation and its linear approximation, \mbox{LinDistFlow}. We focus on a balanced single-phase equivalent model to present the key concepts and then extend to unbalanced three-phase networks in Section~\ref{sec:Multi-Phase Grids}. 

\subsection{Notation}
\begin{figure}[t!]
\centering
\includegraphics[width=0.49\textwidth]{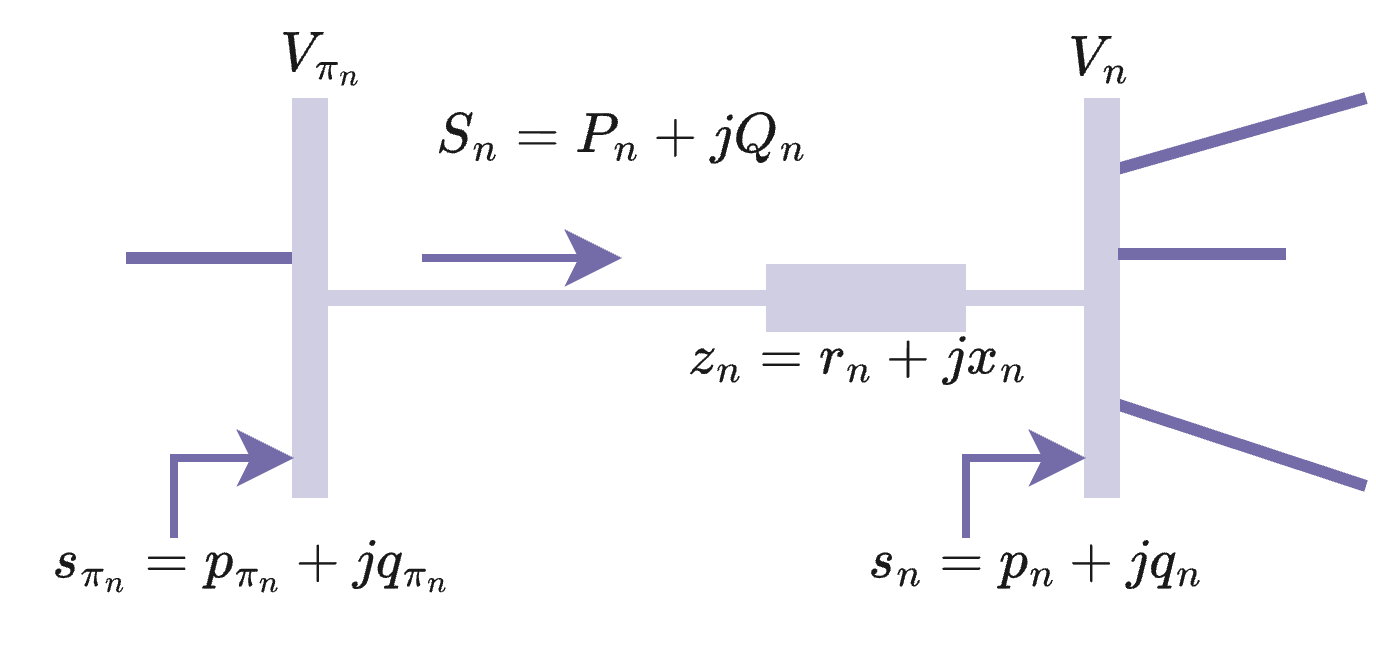}
\caption{A 2-bus system with line $(\pi_n,n)$ feeding bus $n$ from its parent $\pi_{n}$.}
\label{fig:two-bus}
\end{figure}

We first establish notation. Let operators $(\,\cdot\,)^{-1}$, $|\,\cdot\,|$, $(\,\cdot\,)^{\top}$, and $(\,\cdot\,)^{-\top}$ denote the square matrix inverse, the absolute value of a number, the transpose of a matrix/vector, and the transpose of a square matrix inverse, respectively. Let $\mathcal{N}:=\{0, 1, \hdots, n\}$ and $\mathcal{E}:=\{1, \hdots, n\}$ denote the sets of buses and lines, respectively, in a distribution network, where $|\mathcal{E}| = |\mathcal{N}|-1$ for radial networks. Each bus $n \in \mathcal{N}$ has a voltage $V_{n}$. 
\textcolor{black}{The set of all non-substation buses is denoted by $\mathcal{N}^{\prime}:=\{1,2,...,n\}$}.
Each bus $n \in \mathcal{N}^{\prime}$ has a parent (upstream) bus denoted as $\pi_{n}$. Each line $(\pi_{n}, n) \in \mathcal{E}$ (connection between bus $n$ and its parent $\pi_{n}$) has impedance $z_{n}=r_{n}+\mathrm{j} x_{n}$ with resistance $r_{n}$ and reactance $x_{n}$, where $\mathrm{j}:=\sqrt{-1}$ (note that lines are identified by their child buses to simplify notation; see Fig.~\ref{fig:two-bus}). Let $I_{n}$ and $S_{n}= P_{n}+\mathrm{j}Q_{n}$ denote the current and complex power flows, respectively, on line $(\pi_{n}, n)$. The net power injection at bus $n$ is $s_{n}=p_{n}+jq_{n}$. Additional variables include squared voltage and current magnitudes $v_{n}=|V_{n}|^{2}$ and $\ell_{n}=|I_{n}|^{2}$. $\mathbf{D}_{r}=\text{diag}(\mathbf{r})$ and $\mathbf{D}_{x}=\text{diag}(\mathbf{x})$ are diagonal matrices of resistances and reactances, and $\mathbf{p}=[p_{1}, \hdots, p_{n}]^{\top}$, $\mathbf{q}=[q_{1}, \hdots, q_{n}]^{\top}$, $\mathbf{P}=[P_{1}, \hdots, P_{n}]^{\top}$, and $\mathbf{Q}=[Q_{1},\hdots, Q_{n}]^{\top}$ are vectors of active/reactive power injections and flows. Define $\hat{\mathbf{A}}= [\mathbf{a}_{0}~\mathbf{A}]$ as the $|\mathcal{E}| \times |\mathcal{N}|$ branch-bus incidence matrix describing the connections between the system's buses and branches, where $\mathbf{a}_{0}$ is the length-$|\mathcal{E}|$ vector associated with the substation bus and $\mathbf{A}$ is the reduced branch-bus matrix for all buses other than the substation bus. Let $\mathbf{v}:=[|V_{1}|^{2}, \hdots, |V_{n}|^{2}]^{\top}$ represent squared voltage magnitudes and $v_{0}= |V_{0}|^{2}$ correspond to the substation bus. 
\textcolor{black}{To enhance voltage approximation accuracy, our algorithm also optimizes a set of bias parameters. These parameters are formulated as vectors of length $|\mathcal{N}^{\prime}|$ and introduce corrective offsets to the model: $\gamma$ is an offset for the squared voltage magnitudes ($v$), $\rho$ is an offset for the active power injections ($p$), and $\varrho$ is an offset for the reactive power injections ($q$).}

\subsection{DistFlow Model for Single-Phase Equivalent Networks}
\label{subsec:DistFlow}
The \mbox{DistFlow} model~\cite{baran1989network,baran1989optimal1,baran1989optimal2} accurately represents voltage, current, and power flow relationships in radial distribution networks.
The \mbox{DistFlow} model is:
%
\begin{subequations}
\label{eq:DistFlow}
    \begin{align}
        &v_{n} = |V_{n}|^{2}, \quad \forall n\in\mathcal{N} ; \quad \ell_{n} = |I_{n}|^{2},  & \forall n\in\mathcal{N}^{\prime}, \label{eq:DistFlow1}\\
        &\sum_{k: n \rightarrow k} P_{k} = p_{n} + P_{n} - r_{n} \ell_{n}, & \forall n\in\mathcal{N}^{\prime},\label{eq:DistFlow2}\\
        &\sum_{k: n \rightarrow k} Q_{k} = q_{n} + Q_{n} - x_{n} \ell_{n}, & \forall n\in\mathcal{N}^{\prime},\label{eq:DistFlow3}\\
        &v_{n} = v_{\pi_{n}} - 2(r_{n} P_{n} + x_{n} Q_{n}) + (r_{n}^{2} + x_{n}^{2}) \ell_{n}, \hspace*{-0.99em} & \forall n\in\mathcal{N}^{\prime}, \label{eq:DistFlow5}\\
        & v_{\pi_{n}} \ell_n = P_n^2 + Q_n^2, & \forall n\in\mathcal{N}^{\prime}.\label{eq:DistFlow6}
    \end{align}
\end{subequations}
%
Equation~\eqref{eq:DistFlow1} defines variables for the squared voltage magnitudes $v_n$ and squared current flow magnitudes $\ell_n$, which are used in place of the voltage phasors $V_n$ and current phasors~$I_n$. Equations~\eqref{eq:DistFlow2} and~\eqref{eq:DistFlow3} correspond to active and reactive power balance at each bus. Equation~\eqref{eq:DistFlow5} models the change in squared voltage magnitudes across lines. Equation~\eqref{eq:DistFlow6} uses the definition of apparent power to relate the squared voltages $v_n$, squared currents $\ell_n$, and squared apparent power $P_n^2 + Q_n^2$. The \mbox{DistFlow} model is nonlinear due to~\eqref{eq:DistFlow6}. \textcolor{black}{Note that the DistFlow model~\eqref{eq:DistFlow} is equivalent to bus-injection AC power flow models for radial systems~\cite{baran1989optimal1,baran1989optimal2,baran1989network,low2014,bose2015}.}

\subsection{Traditional \mbox{LinDistFlow} Approximation}
\label{subsec:LinDistFlow}
The \mbox{LinDistFlow} approximation linearizes the \mbox{DistFlow} equations by neglecting the line loss terms $r_n \ell_{n}$ in~\eqref{eq:DistFlow2}, $x_n \ell_{n}$ in~\eqref{eq:DistFlow3}, and $(r_{n}^{2} + x_{n}^{2}) \ell_{n}$ in~\eqref{eq:DistFlow5}~\cite{baran1989network,baran1989optimal1,baran1989optimal2}. Without these, the nonlinear equation~\eqref{eq:DistFlow6} can be dropped, with the remaining equations linearly relating the squared voltage magnitudes, the power injections, and the line flows:
%
%
\begin{subequations}
\label{eq:LinDistFlow}
    \begin{align}
        &\sum_{k: n \rightarrow k} P_{k} \approx p_{n} + P_{n}, &\forall n\in\mathcal{N}^{\prime}, \label{eq:LinDistFlow1}\\
        &\sum_{k: n \rightarrow k} Q_{k} \approx q_{n} + Q_{n}, &\forall n\in\mathcal{N}^{\prime}, \label{eq:LinDistFlow2}\\
        &v_{n} \approx v_{\pi_{n}} - 2(r_{n} P_{n} + x_{n} Q_{n}), &\forall n\in\mathcal{N}^{\prime}. \label{eq:LinDistFlow5}
    \end{align}
\end{subequations}
%
%
%
A matrix form of the \mbox{LinDistFlow} approximation is:
%
%
\begin{subequations}
\label{eq:matrixform}
    \begin{align}
        & \mathbf{D}_{r}= \text{diag}(\mathbf{r}), \quad \mathbf{D}_{x}= \text{diag}(\mathbf{x}), \label{eq:matrixform1}\\
        &\mathbf{p} = \mathbf{A}^{\top} \mathbf{P}, \label{eq:matrixform2}\\
        &\mathbf{q} = \mathbf{A}^{\top} \mathbf{Q}, \label{eq:matrixform3}\\
        & \mathbf{A} \mathbf{v} + v_{0} \mathbf{a}_{0} = 2 \mathbf{D}_{r} \mathbf{P} + 2 \mathbf{D}_{x} \mathbf{Q}, \label{eq:matrixform4}\\
        & \mathbf{v} =  v_{0} \mathbf{1} + 2 \mathbf{A}^{-1}\mathbf{D}_{r} \mathbf{A}^{-\top} \mathbf{p} + 2 \mathbf{A}^{-1}\mathbf{D}_{x} \mathbf{A}^{-\top} \mathbf{q}. \label{eq:matrixform5}
    \end{align}
\end{subequations}
%
%

For radial networks where all lines have positive resistance ($r_{n} \geq 0$) and reactance ($x_{n} \geq 0 $), the \mbox{LinDistFlow} approximation overestimates the voltage magnitudes and underestimates the complex power flows required to supply the loads~\cite{low2014}. 
The following section presents an algorithm to reduce this approximation error by optimizing the \mbox{LinDistFlow} parameters.

%% file: text/proposed_algorithm.tex
\section{Optimized LinDistFlow Approximation (OLDF)}
 \label{sec:proposed algorithm}
 
\textcolor{black}{While the traditional LinDistFlow model derives the matrices $D_r$ and $D_x$ directly from the physical line parameters, our approach treats them as optimizable coefficients. The justification for this is that the LinDistFlow model is an \textit{approximation} of the true nonlinear physics, primarily by neglecting the impacts of line losses. Since the model's structure is simplified, using the exact physical parameters does not guarantee the most accurate linear approximation. Instead, by optimally selecting values for the $\mathbf{D}_r$ and $\mathbf{D}_x$ coefficients, we find the ``effective'' resistance and reactance values that cause the simplified linear model to best match the results of the full nonlinear model over a range of operating conditions. 
We note that conceptually similar approaches for selecting parameter values in power flow approximations are employed in, e.g.,~\cite{misra2018optimal, hohmann2018, muhlpfordt2019optimal, liu2019data, buason2022sample, liu2022data, markovic2022parameterized, jia2023overview, jia2023tutorial1, jia2023tutorial2, taheri2023optimizing, chen2023naps, buason2024adaptive, taheri2024improving, taheri2024acinformed, buason2025sample, rosemberg2025differentiable, chen2025optimally, cho2024datadriven}. In Section~\ref{sec:Numerical experiments}, we numerically compare our optimized LinDistFlow formulation to the relevant alternative LinDistFlow approximations to show that optimizing the parameter values is an effective method for improving LinDistFlow approximation accuracy.}

We first generalize the \mbox{LinDistFlow} approximation by introducing bias parameters $\boldsymbol{\gamma}$, $\boldsymbol{\rho}$, and $\boldsymbol{\varrho}$ that offset the squared voltage magnitudes and active and reactive power injections:
\begin{align}
\label{eq:voltage_final}
    \mathbf{v} &=  v_{0} \mathbf{1} + 2 \mathbf{A}^{-1}\mathbf{D}_{r} \mathbf{A}^{-\top} (\mathbf{p} +  \boldsymbol{\rho})+\nonumber\\
    & \quad \quad \quad \quad \quad \quad \quad \quad  \quad \quad 2 \mathbf{A}^{-1}\mathbf{D}_{x} \mathbf{A}^{-\top} (\mathbf{q} +\boldsymbol{\varrho})+ \boldsymbol{\gamma}.
\end{align}

Appropriate selection of parameter values for $\mathbf{D}_r$, $\mathbf{D}_x$, $\boldsymbol{\gamma}$, $\boldsymbol{\rho}$, and $\boldsymbol{\varrho}$ can significantly improve the \mbox{LinDistFlow} approximation's accuracy.
\textcolor{black}{By adjusting the effective nodal power injections $(\mathbf{p}+\boldsymbol{\rho})$ and $(\mathbf{q}+\boldsymbol{\varrho})$, the parameters $\boldsymbol{\rho}$ and $\boldsymbol{\varrho}$ allow the linear model to implicitly account for the voltage impact of system characteristics not directly captured in the basic LinDistFlow equations, such as line losses.}\footnote{\textcolor{black}{Similar adjustment parameters are often employed to improve the accuracy of DC power flow approximations~\cite{stott2009dc}.}}
We next introduce a machine learning-inspired algorithm which optimizes these parameters to reduce the discrepancy between the voltages output by the \mbox{LinDistFlow} approximation~\eqref{eq:voltage_final} and the \mbox{DistFlow} model~\eqref{eq:DistFlow}.

\begin{figure}[t]
\centerline{\includegraphics[width=0.485\textwidth]{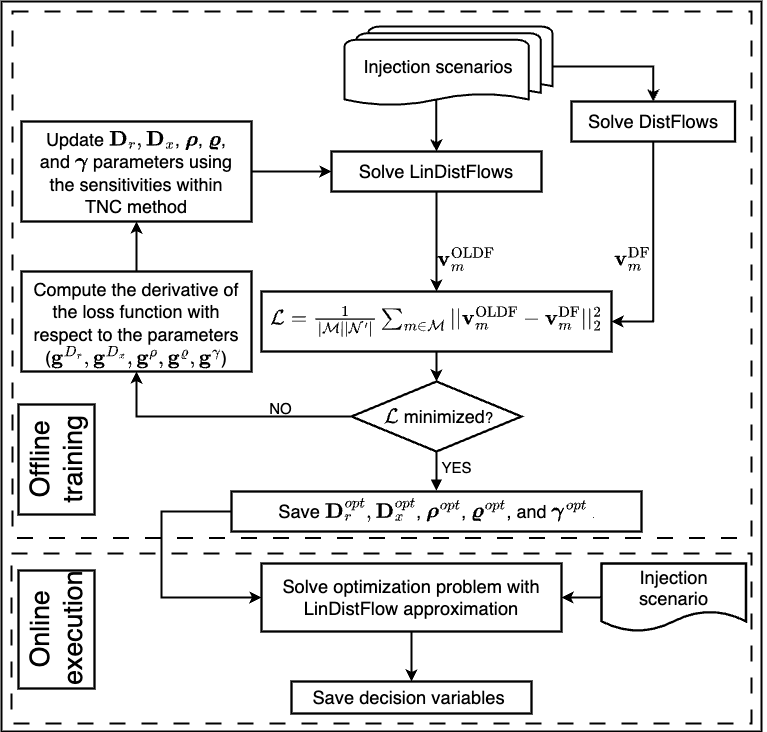}}
\caption{Flowchart depicting the proposed parameter optimization algorithm.}

\label{fig:flowchart}
\end{figure}

As shown in Fig.~\ref{fig:flowchart}, we propose a two-phase algorithm for parameter optimization with an \textit{offline} training phase followed by an \textit{online} application phase. The \textit{offline} phase, conducted once, optimizes the values for the parameters $\mathbf{D}_r$, $\mathbf{D}_x$, $\boldsymbol{\gamma}$, $\boldsymbol{\rho}$, and $\boldsymbol{\varrho}$. This initial computation seeks to align the \mbox{LinDistFlow} approximation with the \mbox{DistFlow} model's behavior across varied operational scenarios. The optimized parameters are then used in the \textit{online} phase to improve accuracy in various applications while enabling tractable solution with linear programming and mixed-integer linear programming solvers.

To optimize the parameters $\mathbf{D}_r$, $\mathbf{D}_x$, $\boldsymbol{\gamma}$, $\boldsymbol{\rho}$, and $\boldsymbol{\varrho}$, Section~\ref{subsec:FormulatingLossFunction} first formulates a loss function that gauges the approximation's accuracy by comparing its voltage predictions to those from the \mbox{DistFlow} model across sampled operational conditions. Section~\ref{subsec:ParameterSensitivityAnalysis} then analytically derives the loss function's sensitivities with respect to the coefficient and bias parameters. Informed by these sensitivities, Section~\ref{sec:ImplementationOptimizationSolution} applies the TNC optimization method to minimize the loss function. 
The resulting optimized parameters are used online for various applications, as benchmarked in Section~\ref{sec:Numerical experiments}.

\subsection{Formulation of the Loss Function}
\label{subsec:FormulatingLossFunction}
We define a loss function, $\mathcal{L}$, as the mean square error between the voltage solutions of the \mbox{DistFlow} model ($\mathbf{v}_{m}^{DF}$) and our optimized \mbox{LinDistFlow} approximation ($\mathbf{v}_{m}^{OLDF}$) across a given set of load scenarios $\mathcal{M} = \{1, 2, \ldots, S\}$:
\begin{align}
\nonumber & \mathcal{L}(\mathbf{D}_r, \mathbf{D}_x, \boldsymbol{\gamma}, \boldsymbol{\rho}, \boldsymbol{\varrho}) = \frac{1}{|\mathcal{M}||\mathcal{N}^{\prime}|}\sum_{m \in \mathcal{M}} ||\mathbf{v}_{m}^{OLDF} - \mathbf{v}_{m}^{DF}||^2_2,\\ %
& \qquad = \frac{1}{|\mathcal{M}||\mathcal{N}^{\prime}|}\sum_{m \in \mathcal{M}} ||v_{0} \mathbf{1} +  2\mathbf{A}^{-1}\mathbf{D}_{r} \mathbf{A}^{-\top} (\mathbf{p} +  \boldsymbol{\rho})+\nonumber\\
& \quad \quad \quad \quad \quad \quad   2 \mathbf{A}^{-1}\mathbf{D}_{x} \mathbf{A}^{-\top} (\mathbf{q} +\boldsymbol{\varrho})+ \boldsymbol{\gamma} - \mathbf{v}_{m}^{DF}||^2_2, \label{eq:objective_function}%
\end{align}%
where normalization by $\frac{1}{|\mathcal{M}||\mathcal{N}^{\prime}|}$ adjusts for the system size and number of samples. 
Using mean square error ensures that larger deviations are more heavily penalized, aligning with typical operational priorities where minimizing the most significant errors is often more important than reducing numerous smaller inaccuracies.

The optimal \mbox{LinDistFlow} parameters minimize $\mathcal{L}$:
\begin{equation}
    \minimize_{\mathbf{D}_r, \mathbf{D}_x, \boldsymbol{\gamma}, \boldsymbol{\rho}, \boldsymbol{\varrho}} \mathcal{L}(\mathbf{D}_r, \mathbf{D}_x, \boldsymbol{\gamma}, \boldsymbol{\rho}, \boldsymbol{\varrho}).
    \label{eq:revisedOptimizationProblem}
\end{equation}

\subsection{Parameter Sensitivity Analysis}
\label{subsec:ParameterSensitivityAnalysis}
We use the TNC optimization method to solve~\eqref{eq:revisedOptimizationProblem}. This method relies on the gradients of the loss function with respect to the parameters, which we present next. We start with the sensitivities for the parameters, represented generically by $\mathbf{g}^{\eta}$, where $\eta$ can be any of the parameter sets $\mathbf{D}_r, \mathbf{D}_x, \boldsymbol{\gamma}, \boldsymbol{\rho}$, and $\boldsymbol{\varrho}$. These sensitivities are calculated as follows:
\begin{subequations}
\label{eq:generic_sensitivity}
\begin{equation}
 \mathbf{g}^{\eta}= \frac{2}{|\mathcal{M}||\mathcal{N}^{\prime}|}\sum_{m \in \mathcal{M}} \Big ( \left. \frac{\partial \mathbf{v}^{OLDF}}{\partial \eta} \right|_ {\mathbf{v}^{OLDF}_{m}} \Big)^{\top}\Big(\mathbf{v}_{m}^{OLDF} - \mathbf{v}_{m}^{DF} \Big),
\end{equation}
where $\eta \in \{\mathbf{D}_r, \mathbf{D}_x, \boldsymbol{\gamma}, \boldsymbol{\rho}, \boldsymbol{\varrho}\}$. The partial derivatives are:
\begin{align}
\frac{\partial \mathbf{v}^{OLDF}}{\partial \mathbf{D}_r} &= 2 \mathbf{A}^{-1} \text{diag}\Big(\mathbf{A}^{-\top} (\mathbf{p} + \boldsymbol{\rho}) \Big), \label{eq:coef_sensitivity}\\
\frac{\partial \mathbf{v}^{OLDF}}{\partial \mathbf{D}_x} &= 2 \mathbf{A}^{-1} \text{diag}\Big(\mathbf{A}^{-\top} (\mathbf{q} + \boldsymbol{\varrho}) \Big), \label{eq:coef_sensitivity2}\\
\frac{\partial \mathbf{v}^{OLDF}}{\partial \boldsymbol{\gamma}} &= \mathbf{I}, \label{eq:bias_sensitivity3}
\\
\frac{\partial \mathbf{v}^{OLDF}}{\partial \boldsymbol{\rho}} &= 2\mathbf{A}^{-1}\mathbf{D}_r\mathbf{A}^{-\top}, \label{eq:bias_sensitivity1}
\\
\frac{\partial \mathbf{v}^{OLDF}}{\partial \boldsymbol{\varrho}} &= 2\mathbf{A}^{-1}\mathbf{D}_x\mathbf{A}^{-\top}. \label{eq:bias_sensitivity2}
\end{align}
\end{subequations}
Here, $\mathbf{I}$ is the identity matrix. These sensitivities enable gradient-based methods such as TNC for optimizing the parameters $\mathbf{D}_r$, $\mathbf{D}_x$, $\boldsymbol{\gamma}$, $\boldsymbol{\rho}$, and $\boldsymbol{\varrho}$, as we will describe next.

\subsection{Implementation of the Optimization Solution}
\label{sec:ImplementationOptimizationSolution}
The gradients in Section~\ref{subsec:ParameterSensitivityAnalysis} enable the application of gradient-based optimization methods such as TNC~\cite{ nocedal2006numerical, nash1984newton} to solve the parameter optimization problem~\eqref{eq:revisedOptimizationProblem}. The choice of TNC as opposed to other gradient-based optimization methods such as Broyden-Fletcher-Goldfarb-Shanno (BFGS) and limited-memory BFGS (\mbox{L-BFGS})~\cite{nocedal2006numerical, nash1984newton} is based on the TNC method's superior scalability in our empirical testing. 
\textcolor{black}{Specifically, for our parameter optimization problem, TNC demonstrated superior practical performance over L-BFGS variants, converging with less computational time, particularly for larger test cases. This advantage in handling large-scale systems is consistent with our findings in related work~\cite{taheri2023optimizing}, where TNC also proved more effective.}

As illustrated in Algorithm~\ref{alg:truncated_newton_with_preconditioning}, TNC iteratively uses Hessian-vector products to approximate the Newton direction, effectively managing memory and computation even in high-dimensional spaces. TNC operates by approximating the solution to Newton's equations for a function's local minimum, truncating early to conserve computational resources while still moving significantly towards the minimum. This is particularly advantageous for problems where evaluating the full Hessian matrix is difficult. To compute~\eqref{eq:revisedOptimizationProblem} in order to find the optimal LinDistFlow parameters, we utilize SciPy's \texttt{scipy.optimize.minimize} TNC implementation with the sensitivities from Section~\ref{subsec:ParameterSensitivityAnalysis}, using gradient norm thresholds and iteration limits as termination criteria.
\textcolor{black}{SciPy's TNC internally computes the required Hessian-vector products ($\mathbf{H}\mathbf{p}$ in Algorithm~\ref{alg:truncated_newton_with_preconditioning}) using finite differences of the analytical gradients from Section~\ref{subsec:ParameterSensitivityAnalysis}, as no custom Hessian-vector product function was supplied. We employed default SciPy TNC values for the Wolfe line search parameters ($\alpha_1$ and $\alpha_2$ in Algorithm~\ref{alg:truncated_newton_with_preconditioning}) and applied no custom preconditioner, effectively treating $\mathbf{M}$ in Algorithm~\ref{alg:truncated_newton_with_preconditioning} as the identity matrix.}

\begin{algorithm}[!th]
\label{alg:truncated_newton_with_preconditioning}
\smaller 

\DontPrintSemicolon
\caption{Truncated Newton (TNC) Method}

\KwInput{\hspace{0cm}$\mathbf{x}_0 = [\mathbf{D}^{\top}_{r}, \mathbf{D}^{\top}_{x}, \boldsymbol{\gamma}^{\top}, \boldsymbol{\rho}^{\top},\boldsymbol{\varrho}^{\top}]^{\top}$: Initial guess\\
\hspace{1cm}$\epsilon$: Tolerance for convergence \\

\hspace{1cm}$\textit{max\_iter}$: Maximum iterations\\
\hspace{1cm}$\mathcal{L}(\mathbf{x}_{k})$: Loss function\\
\hspace{1cm}$\nabla \mathcal{L}(\mathbf{x}_{k})$$=\mathbf{g} = [{\mathbf{g}^{D_r}}^{\top},{\mathbf{g}^{D_x}}^{\top}, {\mathbf{g}^{\gamma}}^{\top}, {\mathbf{g}^{\rho}}^{\top}, {\mathbf{g}^{\varrho}}^{\top}]^{\top}$\\
\hspace{1cm}$\mathbf{M}$: Preconditioning matrix (often a diagonal matrix) \\
\hspace{1cm}$\mathbf{H}$: Hessian or its approximation function \\
\hspace{1cm}$\alpha_1$: Armijo condition constant, small (e.g., $10^{-4}$) \\
\hspace{1cm}$\alpha_2$: Curvature condition constant, between $\alpha_1$ and $1$ 
}

\KwOutput{
    Optimized parameters $\mathbf{x}^*$
}

Initialize $\mathbf{x}_k \leftarrow \mathbf{x}_0$\;
$k \leftarrow 0$\;

\While{$k \leq \textit{max\_iter}$ and $\|\nabla \mathcal{L}(\mathbf{x}_{k})\| > \epsilon$}
{

    $\mathbf{g} \gets \nabla \mathcal{L}(\mathbf{x}_{k})$  \quad\tcp*{Compute the gradient at $\mathbf{x}_{k}$}

    $\mathbf{z} \gets \mathbf{M}^{-1} \mathbf{g}$   \quad \tcp*{Apply the preconditioner}

    $\mathbf{r} \gets -\mathbf{g}$ , $\mathbf{p} \gets \mathbf{z}$

    $\rho_{old} \gets \mathbf{r}^{\top} \mathbf{z}$\;

    \While{$\|\mathbf{r}\| > \epsilon$}{
        $\mathbf{q} \gets \mathbf{H} \mathbf{p}$\qquad  \tcp*{Hessian-vector product}
        $\alpha \gets \frac{\rho_{old}}{\mathbf{p}^{\top} \mathbf{q}}$\;
        $\mathbf{p} \gets \mathbf{z} + \eta \mathbf{p}$\;
        $\mathbf{r} \gets \mathbf{r} - \alpha \mathbf{q}$\;
        $\mathbf{z} \gets \mathbf{M}^{-1} \mathbf{r}$\;
        $\rho_{new} \gets \mathbf{r}^{\top} \mathbf{z}$\;
        $\eta \gets \frac{\rho_{new}}{\rho_{old}}$\;
        $\rho_{old} \gets \rho_{new}$\;
    }


   \tcp*{Wolfe Line Search to determine $\beta$}
    $\beta \gets 1$ \qquad \tcp{Initial step length}
    \While{True}{
        \If{$\mathcal{L}(\mathbf{x}_{k} + \beta \mathbf{p}) \leq \mathcal{L}(\mathbf{x}_{k}) + \alpha_1 \beta \mathbf{g}^{\top} \mathbf{p}$ \textbf{and} $\|\nabla \mathcal{L}(\mathbf{x}_{k} + \beta \mathbf{p})^{\top} \mathbf{p}\| \leq \alpha_2 \|\mathbf{g}^{\top} \mathbf{p}\|$}{
            Break\;
        }
        $\beta \gets \beta / 2$\; 
    }
    $\mathbf{x}_{k+1} \gets \mathbf{x}_{k} + \beta \mathbf{p}$ \tcp*{Update the parameter vector}
    $k \gets k + 1$\;

}

$\mathbf{x}^{*} \leftarrow \mathbf{x}_{k}$ \; 
\end{algorithm}

\section{Three-Phase LinDistFlow (LinDist3Flow)}
\label{sec:Multi-Phase Grids}

We next discuss extensions to unbalanced three-phase distribution systems~\cite{gan2014convex}. The equivalent of \eqref{eq:voltage_final} for LinDist3Flow~is (see Appendix~\ref{appendix:three-phase_derivations} for details):

\begin{align}
\label{eq:voltage_final_3phase}
    \mathbb{V} &=  \mathbb{V}_{0} \mathbf{1} + \mathbb{A}_{3}^{-1}\text{bdiag}(\mathbb{H}^{P}) \mathbb{A}_{3}^{-\top} (\mathbb{P} +  \boldsymbol{\rho}_{3}) + \nonumber\\
    & \quad \quad \quad \quad \quad \mathbb{A}_{3}^{-1}\text{bdiag}(\mathbb{H}^{Q}) \mathbb{A}_{3}^{-\top} (\mathbb{Q} + \boldsymbol{\varrho}_{3}) + \boldsymbol{\gamma}_{3}.
\end{align}
where $\mathbb{V} = \begin{bmatrix} v^a_1 & v^b_1 & v^c_1 & \cdots & v^a_n & v^b_n & v^c_n \end{bmatrix}^{\top}$ is the vector of squared voltage magnitudes for each phase $(a, b, c)$, $\mathbb{A}_{3}$ is the network incidence matrix, $\text{bdiag}(\,\cdot\,)$ is the block diagonal operator, $\mathbb{P} = \begin{bmatrix} p^a_1 & p^b_1 & p^c_1 & \cdots & p^a_n & p^b_n & p^c_n \end{bmatrix}^{\top}$ and $\mathbb{Q} = \begin{bmatrix} q^a_1 & q^b_1 & q^c_1 & \cdots & q^a_n & q^b_n & q^c_n \end{bmatrix}^{\top}$ are the active and reactive power injection vectors, and the $\mathbb{H}$ matrices for each line $(i,j)\in\mathcal{E}$ are:
\begin{equation}
    \mathbb{H}_{ij}^P = \begin{bmatrix}
    -2r_{ij}^{aa} & r_{ij}^{ab} - \sqrt{3}x_{ij}^{ab} & r_{ij}^{ac} + \sqrt{3}x_{ij}^{ac} \\
    r_{ij}^{ba} + \sqrt{3}x_{ij}^{ba} & -2r_{ij}^{bb} & r_{ij}^{bc} - \sqrt{3}x_{ij}^{bc} \\
    r_{ij}^{ca} - \sqrt{3}x_{ij}^{ca} & r_{ij}^{cb} + \sqrt{3}x_{ij}^{cb} & -2r_{ij}^{cc}
    \end{bmatrix}, \label{eq:22}
\end{equation}
\begin{equation}
    \mathbb{H}_{ij}^Q = \begin{bmatrix}
    -2x_{ij}^{aa} & x_{ij}^{ab} + \sqrt{3}r_{ij}^{ab} & x_{ij}^{ac} - \sqrt{3}r_{ij}^{ac} \\
    x_{ij}^{ba} - \sqrt{3}r_{ij}^{ba} & -2x_{ij}^{bb} & x_{ij}^{bc} + \sqrt{3}r_{ij}^{bc} \\
    x_{ij}^{ca} + \sqrt{3}r_{ij}^{ca} & x_{ij}^{cb} - \sqrt{3}r_{ij}^{cb} & -2x_{ij}^{cc}
    \end{bmatrix}. \label{eq:23}
\end{equation}
Analogous to the single-phase algorithm, we train the parameters $\mathbb{H}_{ij}^P$, $\mathbb{H}_{ij}^Q$, $\boldsymbol{\rho}_{3}$, $\boldsymbol{\varrho}_{3}$, and $\boldsymbol{\gamma}_{3}$ for three-phase networks. The parameters $\boldsymbol{\rho}_{3}$, $\boldsymbol{\varrho}_{3}$, and $\boldsymbol{\gamma}_{3}$ are the three-phase versions of $\boldsymbol{\rho}$, $\boldsymbol{\varrho}$, and $\boldsymbol{\gamma}$, respectively.
To this end, we rewrite the loss function~\eqref{eq:objective_function} as:
\begin{align}
\nonumber & \mathcal{L}(\mathbb{H}^{P}, \mathbb{H}^{Q}, \boldsymbol{\gamma}_3, \boldsymbol{\rho}_3, \boldsymbol{\varrho}_3) = \frac{1}{|\mathcal{M}||3\mathcal{N}^{\prime}|}\sum_{m \in \mathcal{M}} ||\mathbb{V}_{m}^{OLDF} - \mathbb{V}_{m}^{DF}||^2_2,\\ %
& \quad = \frac{1}{|\mathcal{M}||3\mathcal{N}^{\prime}|}\sum_{m \in \mathcal{M}} ||\mathbb{V}_{0} \mathbf{1} + \mathbb{A}_{3}^{-1}\text{bdiag}(\mathbb{H}^{P}) \mathbb{A}_{3}^{-\top} (\mathbb{P} +  \boldsymbol{\rho}_{3}) + \nonumber\\
    & \quad \quad \quad  \mathbb{A}_{3}^{-1}\text{bdiag}(\mathbb{H}^{Q}) \mathbb{A}_{3}^{-\top} (\mathbb{Q} + \boldsymbol{\varrho}_{3}) + \boldsymbol{\gamma}_{3} - \mathbb{V}_{m}^{DF}||^2_2. \label{eq:objective_function_3}%
\end{align}%
The sensitivity of the loss function \eqref{eq:objective_function_3} with respect to the $\mathbb{H}_{ij}^P$, $\mathbb{H}_{ij}^Q$, $\boldsymbol{\rho}_{3}$, $\boldsymbol{\varrho}_{3}$, and $\boldsymbol{\gamma}_{3}$ parameters, necessary for gradient-based optimization methods such as TNC, can be computed similarly to the single-phase case as in \eqref{eq:generic_sensitivity}. This yields:
\begin{subequations}
\label{eq:generic_sensitivity_3}
\begin{equation}
 \mathbf{g}^{\eta}= \frac{2}{|\mathcal{M}||3\mathcal{N}^{\prime}|}\sum_{m \in \mathcal{M}} \Big(\left. \frac{\partial \mathbb{V}^{OLDF}}{\partial \eta} \right|_ {\mathbb{V}^{OLDF}_{m}}\Big) ^{\top}\Big(\mathbb{V}_{m}^{OLDF} - \mathbb{V}_{m}^{DF} \Big),
\end{equation}
where $\eta \in \{\text{bdiag}(\mathbb{H}^{P}), \text{bdiag}(\mathbb{H}^{Q}), \boldsymbol{\gamma}_3, \boldsymbol{\rho}_3, \boldsymbol{\varrho}_3\}$, i.e., the parameters to be optimized. The partial derivatives are:
\begin{align}
\frac{\partial \mathbb{V}^{OLDF}}{\partial \text{bdiag}(\mathbb{H}^{P})} &= \Big(\mathbb{A}_3^{-\top} (\mathbb{P} +  \boldsymbol{\rho}_3) \Big)^{\top} \otimes \mathbb{A}_3^{-1}, \label{eq:coef_sensitivity_3}
\\
\frac{\partial \mathbb{V}^{OLDF}}{\partial \text{bdiag}(\mathbb{H}^{Q})} &=  \Big(\mathbb{A}_3^{-\top} (\mathbb{Q} +  \boldsymbol{\varrho}_3) \Big)^{\top} \otimes \mathbb{A}_3^{-1}, \label{eq:coef_sensitivity2_3}
\\
\frac{\partial \mathbb{V}^{OLDF}}{\partial \boldsymbol{\gamma}_3} &= \mathbf{I}, \label{eq:bias_sensitivity3_3}
\\
\frac{\partial \mathbb{V}^{OLDF}}{\partial \boldsymbol{\rho}_3} &=\mathbb{A}_3^{-1}\text{bdiag}(\mathbb{H}^{P})\mathbb{A}_3^{-\top}, \label{eq:bias_sensitivity1_3}
\\
\frac{\partial \mathbb{V}^{OLDF}}{\partial \boldsymbol{\varrho}_3} &=\mathbb{A}_3^{-1}\text{bdiag}(\mathbb{H}^{Q})\mathbb{A}_3^{-\top}. \label{eq:bias_sensitivity2_3}
\end{align}
\end{subequations}
Here, \(\otimes\) denotes the Kronecker product.

%% file: text/results.tex
\section{Numerical Analysis}
\label{sec:Numerical experiments}
This section empirically benchmarks the proposed algorithm  ``optimized LinDistFlow'' (OLDF) against several other related power flow linearizations. Specifically, in addition to the nonlinear \mbox{DistFlow} model~\eqref{eq:DistFlow} that provides the ground truth via an AC power flow solution, we also benchmark our optimized \mbox{LinDistFlow} against the traditional \mbox{LinDistFlow} approximation (LDF)~\cite{baran1989optimal1,baran1989optimal2,baran1989network}, the parameterized linear power flow (PLPF) approximation from~\cite{markovic2022parameterized}, the Lossy \mbox{DistFlow} (LoDF) approximation from~\cite{schweitzer2020lossy}, and the decoupled linear power flow (DLPF) approximation from~\cite{yang2017linearized}. We use various test distribution systems and loading scenarios to replicate the benchmarking methodologies adopted in this literature. These include the balanced single-phase equivalent test cases \texttt{IEEE 33-bus}, \texttt{IEEE 69-bus}, and modified \texttt{IEEE 123-bus} from~\cite{bolognani2015existence} as well as the \texttt{22-bus}, \texttt{85-bus}, \texttt{141-bus}, \texttt{case533mt-hi}, and \texttt{case-eu906} systems from M{\sc atpower}~\cite{matpower}. For three-phase distribution networks, we use the \texttt{IEEE 13-bus}, \texttt{IEEE 37-bus}, and \texttt{IEEE 123-bus} test cases.

\subsection{Algorithm Training}
\label{subsec:Algorithm Training}
\begin{figure}[ht]
\centering
\includegraphics[width=0.48\textwidth]{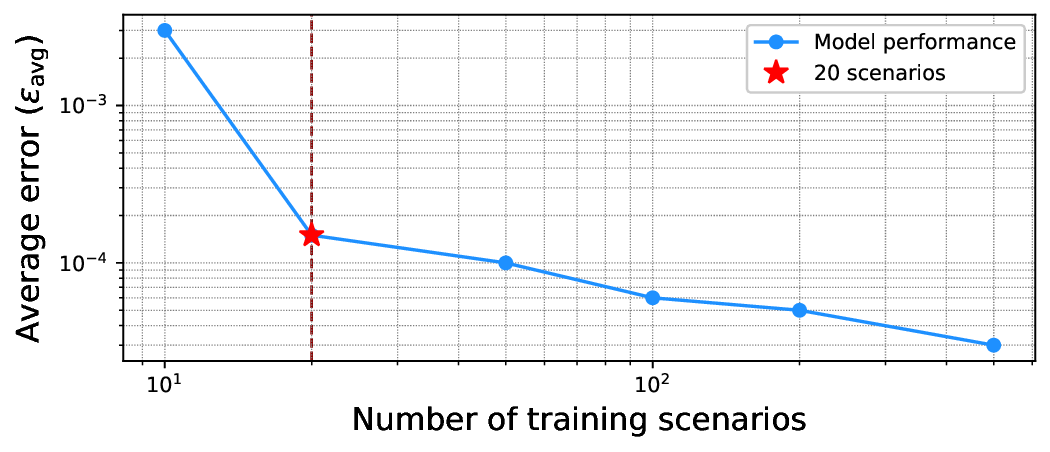} 
\caption{Average error ($\epsilon_{\text{avg}}^{OLDF}$) as a function of the number of training scenarios for the \texttt{IEEE 33-bus} test system.}

\label{fig:sample_sensitivity_plot}
\end{figure}

As in~\cite{markovic2022parameterized}, we use $20$ power injection scenarios during our algorithm's training phase. \textcolor{black}{This selection is supported by a sensitivity analysis (see Fig.~\ref{fig:sample_sensitivity_plot}), which showed diminishing returns in accuracy improvement with a larger number of scenarios, yielding a balance between model fidelity and training efficiency.}  These scenarios scale the nominal power injections at each bus by normally distributed multipliers with a mean of one and standard deviation of $35\%$. For the nonlinear \mbox{DistFlow} solutions, we use PowerModels.jl~\cite{coffrin2018powermodels} and OpenDSS with the OpenDSSDirect.py package for single- and three-phase networks, respectively, on a PACE computing node at Georgia Tech with a 24-core CPU and 32~GB of RAM. The training algorithm is implemented in Python 3 in a Jupyter Notebook using the TNC method from scipy.optimize.minimize with objective function \eqref{eq:objective_function}, Jacobian $\mathbf{g} = [{\mathbf{g}^{D_r}}^{\top},{\mathbf{g}^{D_x}}^{\top}, {\mathbf{g}^{\gamma}}^{\top}, {\mathbf{g}^{\rho}}^{\top}, {\mathbf{g}^{\varrho}}^{\top}]^{\top}$, convergence tolerance of $1\times10^{-6}$ per unit, and iteration limit of $100$.\footnote{Code is available at \url{https://github.com/BabakTaheri1/OLDF}}

\subsection{Performance Metrics}
\label{subsec:Performance Metrics}
We quantify approximation accuracy by comparing the voltage magnitude outputs (denoted $\mathbf{v}^{[\text{model}]}$, where $[\text{model}]$ is OLDF, PLPF, LoDF, LDF, or DLPF for the various approximations) against the nonlinear \mbox{DistFlow} solutions (denoted $\mathbf{v}^{DF}$). We use max and mean error metrics in per unit (p.u.):
\begin{align}
\varepsilon_{\text{max}}^{[\text{model}]} &= \max_{m \in \mathcal{M}} \|\mathbf{v}_{m}^{[\text{model}]}-\mathbf{v}_{m}^{\text{DF}}\|_{\infty} \\
\varepsilon_{\text{avg}}^{[\text{model}]} &= \frac{1}{|\mathcal{M}||\mathcal{N}^{\prime}|} \sum_{m \in \mathcal{M}} \|\mathbf{v}_{m}^{[\text{model}]}-\mathbf{v}_{m}^{\text{DF}}\|_{1}
\end{align}
where $|\mathcal{M}|$ is the number of testing samples, $|\mathcal{N}^{\prime}|$ is the number of non-root nodes in the distribution systems, $\|\cdot\|_{\infty}$ is the $L_{\infty}$-norm, and $\|\cdot\|_{1}$ is the $L_{1}$-norm.

\begin{figure*}[t!]
\centering
\subfloat[\small ]{
\includegraphics[width=0.485\textwidth]{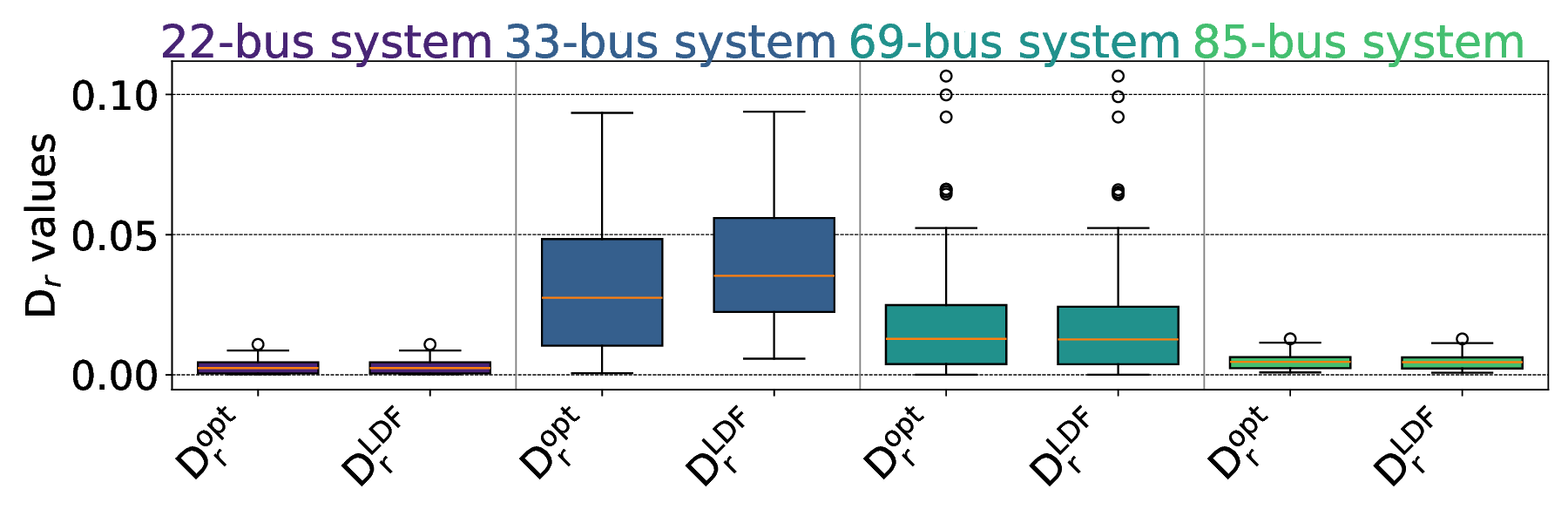}
    \label{fig:Dr-box-plot}
}
\hfill
\subfloat[\small  ]{
\includegraphics[width=0.485\textwidth]{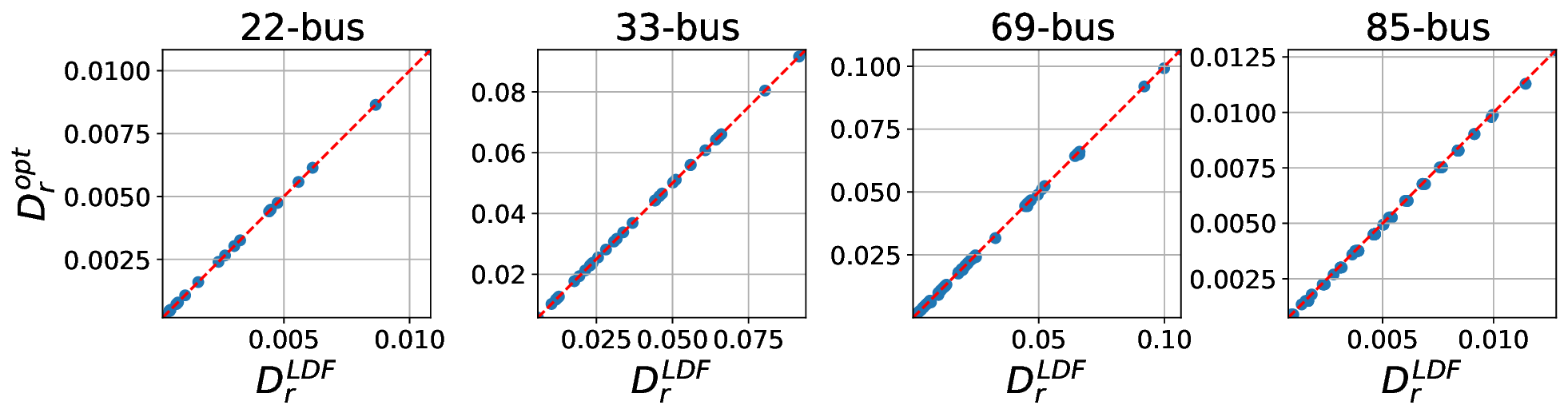}
    \label{fig:Dr-scatter}
}
\caption{(a) Boxplots showing the distributions of the $\mathbf{D}_r$ parameter values for multiple test cases. Each test case is represented by two boxplots indicating the traditional and optimal $\mathbf{D}_r$ parameter values. (b) Scatter plots comparing the coefficient values  $\mathbf{D}_r^{LDF}$ and $\mathbf{D}_r^{opt}$ for various test cases.}
\label{fig:Dr}
\end{figure*}
\begin{figure*}[t!]
\centering
\subfloat[\small ]{
\includegraphics[width=0.485\textwidth]
{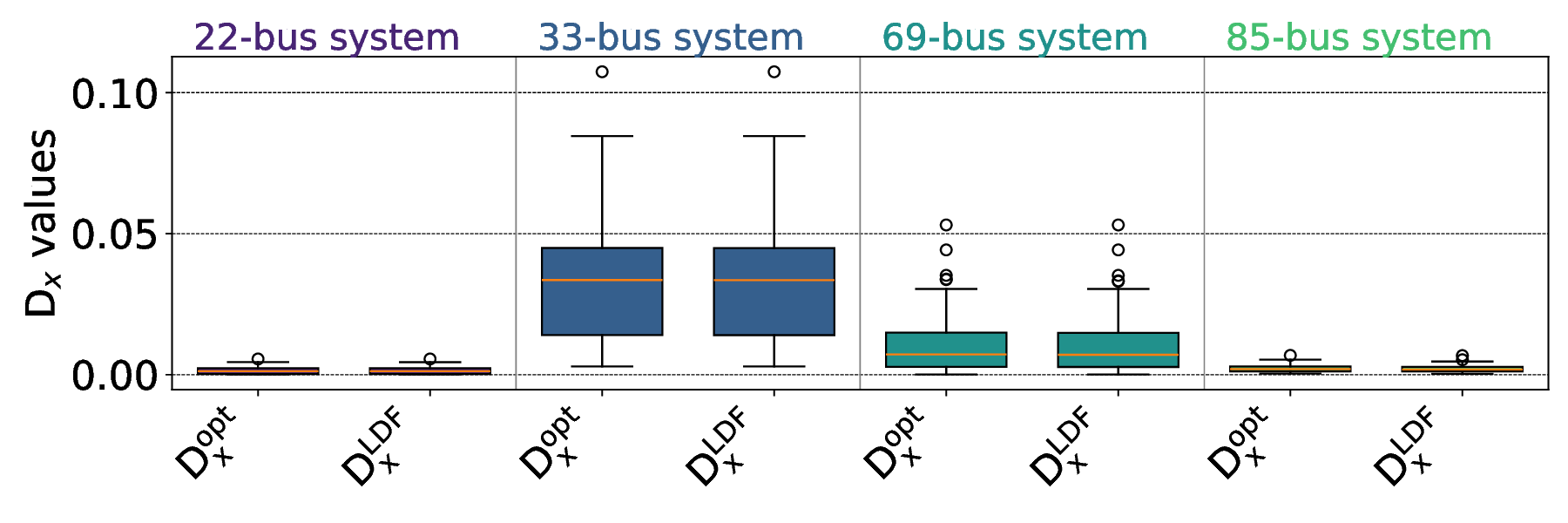}
    \label{fig:Dx-box-plot}
}
\hfill
\subfloat[\small  ]{
\includegraphics[width=0.485\textwidth]{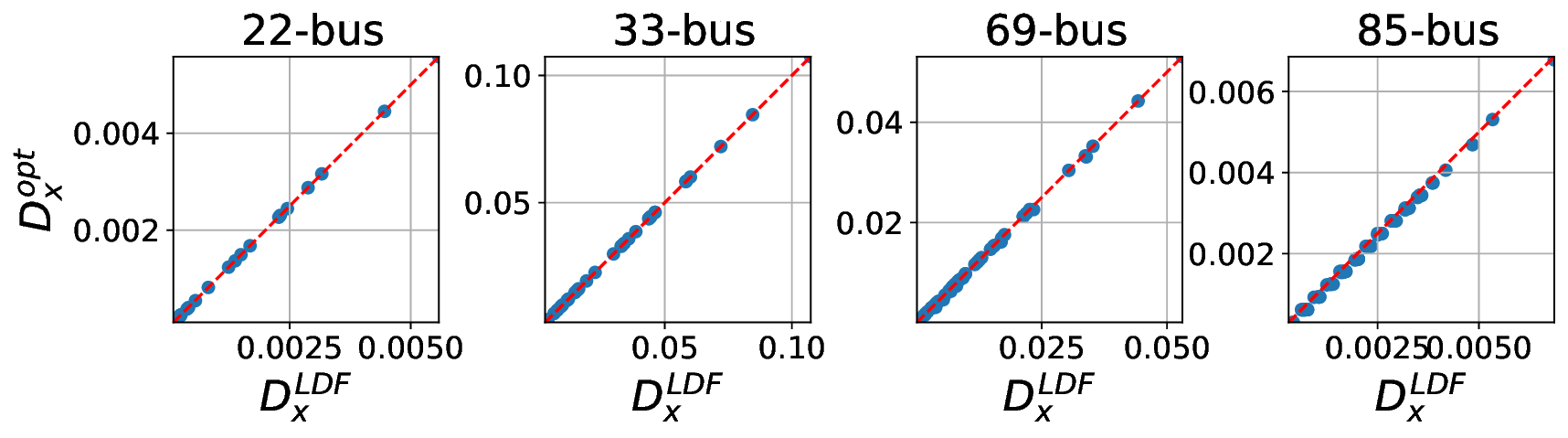}
    \label{fig:Dx-scatter}
}
\caption{(a) Boxplots showing the distributions of the $\mathbf{D}_x$ parameter values for multiple test cases. Each test case is represented by two boxplots indicating the traditional and optimal $\mathbf{D}_x$ parameter values. (b) Scatter plots comparing the coefficient values  $\mathbf{D}_x^{LDF}$ and $\mathbf{D}_x^{opt}$ for various test cases.}

\label{fig:Dx}
\end{figure*}

\begin{figure*}[t!]
\centering
\subfloat[\small ]{
\includegraphics[width=0.485\textwidth]{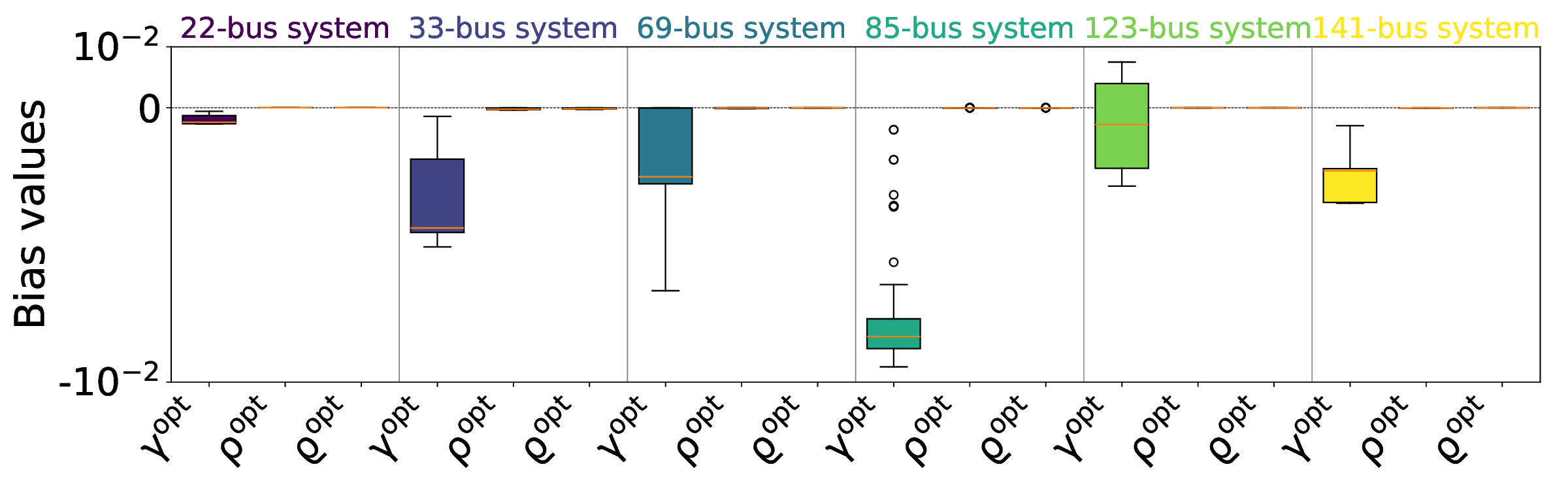}
    \label{fig:bias-boxplot}
}
\hfill
\subfloat[\small  ]{
\includegraphics[width=0.485\textwidth]{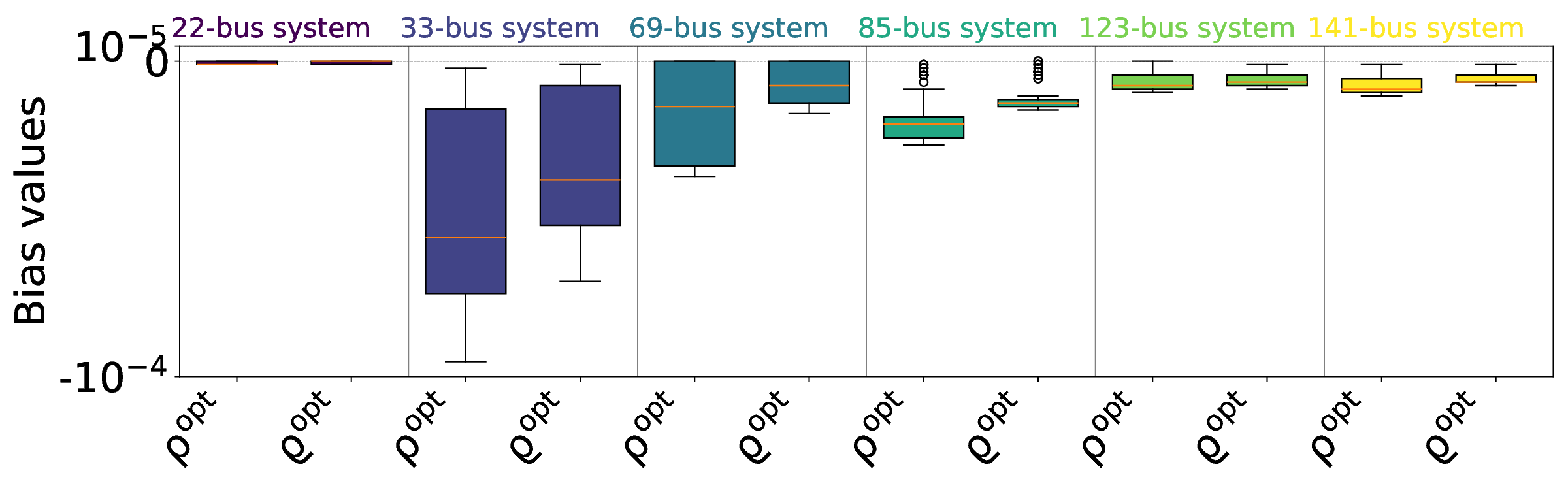}
    \label{fig:rho-boxplot}
}
\caption{Boxplots showing the distributions of the bias parameters for multiple test cases. Each test case is represented by three boxplots indicating the $\boldsymbol{\gamma}^{opt}$, $\boldsymbol{\rho}^{opt}$, and $\boldsymbol{\varrho}^{opt}$ parameter values. (b) Plotting only $\boldsymbol{\rho}^{opt}$ and $\boldsymbol{\varrho}^{opt}$ parameters for better comparison.}

\label{fig:biasboxplot}
\end{figure*}
\begin{figure*}[t!]
\centering
\subfloat[\small \texttt{37-bus} ]{
\includegraphics[width=0.485\textwidth]{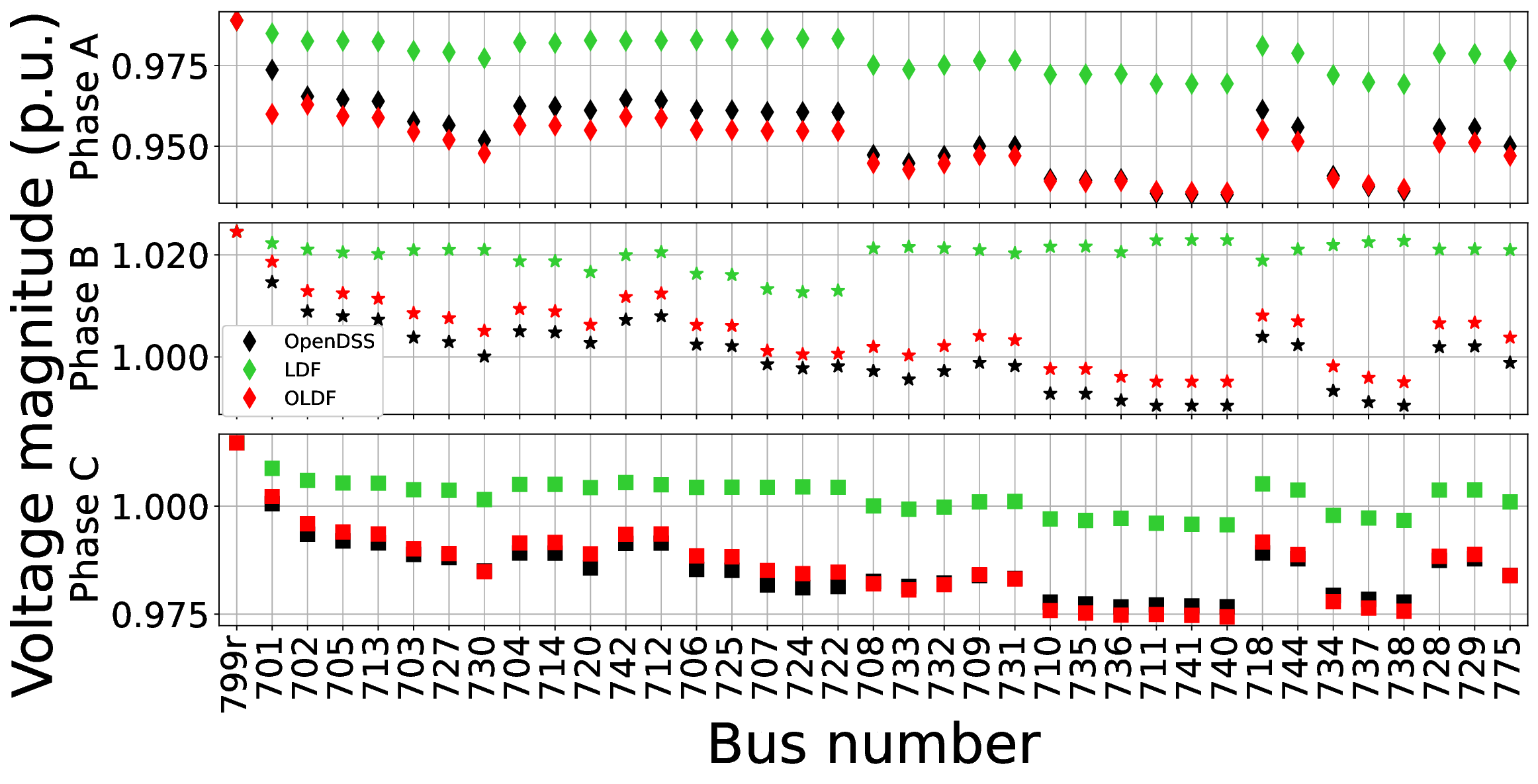}
\label{fig:85}
}
\hfill
\subfloat[\small Modified \texttt{123-bus} ]{
\includegraphics[width=0.485\textwidth]{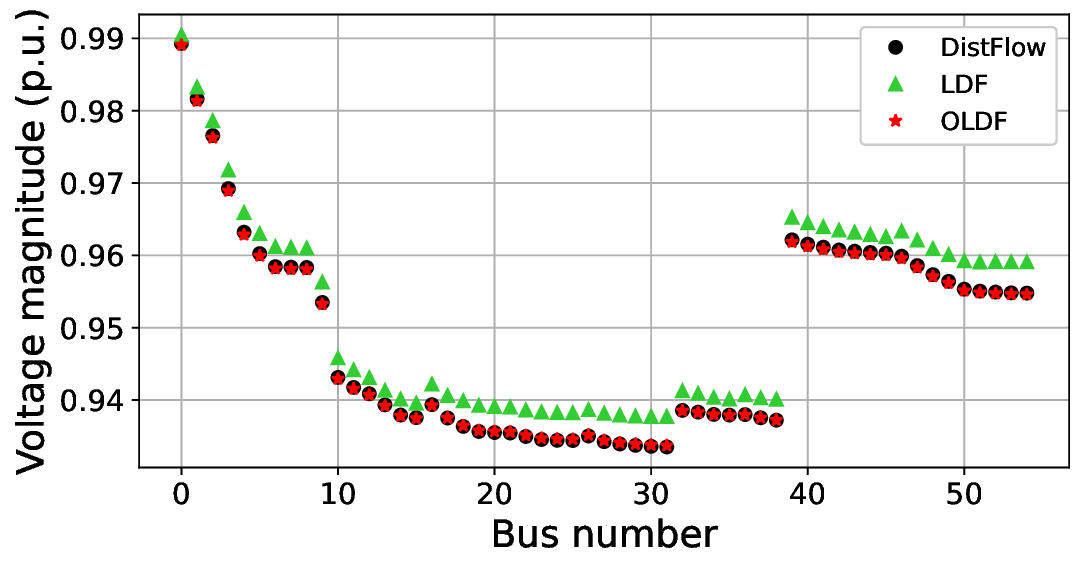}
\label{fig:123}
}
\caption{Voltage profiles for the base case using the nonlinear DistFlow (OpenDSS in black), LinDistFlow with initial parameters (LDF in green), and \textcolor{black}{proposed OLDF (LinDistFlow with optimized parameters in red)}.}

\label{fig:testing}
\end{figure*}

\begin{table*}[t]
    \centering
    \caption{Model Evaluation - Base Load}
    \renewcommand{\arraystretch}{1.3}
    \label{table:Model evaluation- base load}
    \begin{tabular}{|p{0.055cm}>{\centering\arraybackslash}c|ccccc|ccccc|}
    \hline
    &\text{Test case} & \(\varepsilon_{avg}^{LDF}\) & \(\varepsilon_{avg}^{PLPF}\) & \(\varepsilon_{avg}^{LoDF}\) & \(\varepsilon_{avg}^{DLPF}\) & \(\varepsilon_{avg}^{OLDF}\) & \(\varepsilon_{\max}^{LDF}\) & \(\varepsilon_{\max}^{PLPF}\) & \(\varepsilon_{\max}^{LoDF}\) & \(\varepsilon_{\max}^{DLPF}\) & \(\varepsilon_{\max}^{OLDF}\) \\
    \hline \hline
    \multirow{7}{*}{\rotatebox{90}{\textbf{\underline{$1\phi$}}}}
    & 22-bus & 0.00023 & 0.00014 & 0.00236 & 0.00040 & \(\mathbf{0.00001}\) & 0.00030 & 0.00025 & 0.00314 & 0.00066 & \(\mathbf{0.00001}\) \\
    & 33-bus & 0.00198 & 0.00080 & 0.00288 & 0.00368 & \(\mathbf{0.00015}\) & 0.00284 & 0.00125 & 0.00402 & 0.00638 & \(\mathbf{0.00025}\) \\
    & 69-bus & 0.00119 & 0.00075 & 0.00112 & 0.00186 & \(\mathbf{0.00023}\) & 0.00388 & 0.00290 & 0.00327 & 0.00766 & \(\mathbf{0.00094}\) \\
    & 85-bus & 0.00531 & 0.00180 & 0.00206 & 0.00942 & \(\mathbf{0.00056}\) & 0.00663 & 0.00221 & 0.00261 & 0.01377 & \(\mathbf{0.00075}\) \\
    & 123-bus & 0.00218 & 0.00160 & 0.00494 & 0.00348 & \(\mathbf{0.00018}\) & 0.00255 & 0.00186 & 0.00579 & 0.00460 & \(\mathbf{0.00034}\) \\
    & 141-bus & 0.00152 & 0.00071 & 0.00326 & 0.00280 & \(\mathbf{0.00003}\) & 0.00207 & 0.00099 & 0.00409 & 0.00453 & \(\mathbf{0.00004}\) \\
    & 533-bus & 0.00031 & - & - & - & \(\mathbf{0.00001}\) & 0.00079 & - & - & - & \(\mathbf{0.00005}\) \\
    &906-bus & 0.02552 & - & - & - & \(\mathbf{0.00037}\) &0.02566  & - & - & - & \(\mathbf{0.00119}\) \\
    \hdashline
    \multirow{3}{*}{\rotatebox{90}{\textbf{\underline{$3\phi$}}}}
    & 13-bus & 0.02642 & - & - & - & \(\mathbf{0.00901}\) & 0.05899 & - & - & - & \(\mathbf{0.03558}\) \\
    & 37-bus & 0.02022 & - & - & - & \(\mathbf{0.00325}\) & 0.03435 & - & - & - & \(\mathbf{0.00710}\) \\
    & 123-bus & 0.01388 & - & - & - & \(\mathbf{0.00340}\) & 0.03707 & - & - & - & \(\mathbf{0.01322}\) \\
    \hline
    \end{tabular}
    \begin{tablenotes}
    \item[*] \scriptsize \hspace{4em}$1\phi$ stands for balanced single-phase equivalent networks, and $3\phi$ stands for unbalanced three-phase networks.
    \item[*] \scriptsize \hspace{4em}The best performing method (smallest loss function) is bolded for each test case. All values are in per unit.
    \end{tablenotes}
\end{table*}

\begin{table*}[t]
    \centering
    \caption{Model Evaluation - High Load}
    \renewcommand{\arraystretch}{1.2}
    \label{tab:comparison_High}
    \begin{tabular}{|p{0.055cm}>{\centering\arraybackslash}c|ccccc|ccccc|}
    \hline
&\text{Test case} & \(\varepsilon_{avg}^{LDF}\) & \(\varepsilon_{avg}^{PLPF}\) & \(\varepsilon_{avg}^{LoDF}\) & \(\varepsilon_{avg}^{DLPF}\) & \(\varepsilon_{avg}^{OLDF}\) & \(\varepsilon_{\max}^{LDF}\) & \(\varepsilon_{\max}^{PLPF}\) & \(\varepsilon_{\max}^{LoDF}\) & \(\varepsilon_{\max}^{DLPF}\) & \(\varepsilon_{\max}^{OLDF}\) \\
\hline \hline
\multirow{7}{*}{\rotatebox{90}{\textbf{\underline{$1\phi$}}}}
&22-bus & 0.00050 & $\mathbf{0.00020}$ & 0.00376 & 0.00091 & 0.00028 & 0.00132 & $\mathbf{0.00080}$ & 0.00733 & 0.00280 & 0.00097 \\
&33-bus & 0.00418 & 0.00239 & 0.00556 & 0.00795 & $\mathbf{0.00224}$ & 0.01573 & $\mathbf{0.01018}$ & 0.01713 & 0.03133 & 0.01254 \\
&69-bus & 0.00254 & 0.00199 & 0.00278 & 0.00403 & $\mathbf{0.00129}$ & 0.02253 & 0.01975 & 0.01926 & 0.03929 & $\mathbf{0.01572}$ \\
&85-bus & 0.01156 & 0.00817 & 0.01037 & 0.02055 & $\mathbf{0.00719}$ & 0.04700 & 0.03471 & $\mathbf{0.02487}$ & 0.08027 & 0.03373 \\
&123-bus & 0.00464 & 0.00301 & 0.00900 & 0.00555 & $\mathbf{0.00258}$ & 0.01294 & 0.00800 & 0.02017 & 0.02173 & $\mathbf{0.00766}$ \\
&141-bus & 0.00323 & 0.00234 & 0.00601 & 0.00608 & $\mathbf{0.00182}$ & 0.01070 & $\mathbf{0.00648}$ & 0.01501 & 0.02134 & 0.00802 \\
&533-bus & 0.00067 & - & - & - & \(\mathbf{0.00037}\) & 0.00362 & - & - & - & \(\mathbf{0.00278}\) \\
&906-bus &0.02501  & - & - & - & \(\mathbf{0.00358}\) &0.02571  & - & - & - & \(\mathbf{0.00949}\) \\
\hdashline
\multirow{3}{*}{\rotatebox{90}{\textbf{\underline{$3\phi$}}}}
&13-bus &0.06549  & - & - & - & \(\mathbf{0.05147}\) & 0.18626 & - & - & - & \(\mathbf{0.14138}\) \\
&37-bus &0.06294  & - & - & - & \(\mathbf{0.05907}\) &0.22921  & - & - & - & \(\mathbf{0.20931}\) \\
&123-bus&0.03884  & - & - & - & \(\mathbf{0.02379}\) &0.13387  & - & - & - & \(\mathbf{0.12065}\) \\

        \hline
    \end{tabular}
    \begin{tablenotes}
     \item[*] \scriptsize \hspace{4em}$1\phi$ stands for balanced single-phase equivalent networks, and $3\phi$ stands for unbalanced three-phase networks.
    \item[*] \scriptsize  \hspace{4em}The best performing method (smallest loss function) is bolded for each test case. All values are in per unit.
\end{tablenotes}
\end{table*}

\subsection{Parameter Optimization Analysis}
\label{subsec:Parameter Optimization Analysis}
We next present the results of the parameter optimization across various test cases by plotting the parameter values from the traditional \mbox{LinDistFlow} approximation and the \mbox{LinDistFlow} with optimized parameter values.

The box plots in Figs.~\ref{fig:Dr} and~\ref{fig:Dx} show the distributions of $\mathbf{D}_r$ and $\mathbf{D}_x$ parameter values, respectively, for the traditional \mbox{LinDistFlow} and our optimized parameters. Each box plot captures the interquartile range (IQR) with a median line. The whiskers extend to 1.5 times the IQR, with outliers represented as individual points. The horizontal lines at the whiskers' ends indicate the $90^{\textrm{th}}$ percentile.
For each test case, the box plots display two distributions: the \mbox{LinDistFlow} ($\mathbf{D}_{r}^{LDF}$ or $\mathbf{D}_{x}^{LDF}$) and the results from our optimization algorithm ($\mathbf{D}_{r}^{opt}$ and $\mathbf{D}_{x}^{opt}$).
The optimized parameter values' distributions align closely with those from existing heuristics for $\mathbf{D}_r$ and $\mathbf{D}_{x}$. This indicates that our algorithm yields parameter values in a reasonable range, consistent with conventional heuristics.

Additionally, scatter plots accompanying these box plots compare \mbox{LinDistFlow} ($\mathbf{D}_r^{LDF}$, $\mathbf{D}_x^{LDF}$) and optimized ($\mathbf{D}_r^{opt}$, $\mathbf{D}_x^{opt}$) parameter values. The red dashed line at $45^\circ$ in each subplot signifies a one-to-one correlation in the parameter values. 
These plots also show that the optimized parameters are broadly similar to those from existing heuristics, thus aligning with longstanding power engineering intuition that the line resistances and reactances are key parameters in dictating power flows. Despite the overall consistency with traditional \mbox{LinDistFlow} parameter choices, our results show that the optimized parameters yield significant accuracy improvements.

The box plots in Fig.~\ref{fig:biasboxplot} show the distribution of the bias parameters. Illustrated in Fig.~\ref{fig:bias-boxplot}, the optimal values of $\boldsymbol{\rho}^{opt}$ and $\boldsymbol{\varrho}^{opt}$ are notably smaller in magnitude when compared to those of $\boldsymbol{\gamma}^{opt}$. Consequently, Fig.~\ref{fig:rho-boxplot} presents the $\boldsymbol{\rho}^{opt}$ and $\boldsymbol{\varrho}^{opt}$ parameters separately to highlight their distinct distributions. This disparity suggests simplifying the model by focusing on optimizing only three parameters: $\mathbf{D}_{r}$, $\mathbf{D}_{x}$, and $\boldsymbol{\gamma}$. However, our experimental results indicate that this simplification performs poorly for high-load scenarios. We therefore recommend optimizing all of the parameters.


\subsection{Algorithm Testing}
\label{subsec:Algorithm Testing}
We next characterize the optimized parameter values' performance relative to alternative \mbox{LinDistFlow} formulations under three types of load scenarios: \textit{base}, \textit{high}, and \textit{random} load.

\subsubsection{Base Load Evaluation}
\label{subsec:Base Load Evaluation}
We first assess the \mbox{LinDistFlow} approximation accuracy when using the optimized parameter values on the base loading scenarios in the test cases. Table~\ref{table:Model evaluation- base load} shows the OLDF performance metrics, notably the maximum and average voltage magnitude estimation errors ($\varepsilon_{\max}$ and $\varepsilon_{avg}$), compared to the alternative models PLPF, LoDF, LDF, and DLPF. As shown in this table, the OLDF model consistently outperforms its counterparts for all the test cases.

Detailing two examples, Fig.~\ref{fig:testing} shows the voltage profiles for the \texttt{37-bus} and the modified \texttt{123-bus} test cases with the base case loading. While the voltages from the traditional \mbox{LinDistFlow} overestimate the true values, the optimized parameters result in a close alignment with the true \mbox{DistFlow} solution.

\subsubsection{High Load Evaluation}
\label{subsec:High Load Evaluation}
Following the methodology described in~\cite{markovic2022parameterized}, we generated high-load scenarios by scaling the base loads with a factor that ranges from $[-2, -1] \cup [1, 2]$ at a granularity of $\frac{1}{14}$, yielding $30$ distinct test scenarios. As shown in Table~\ref{tab:comparison_High}, our proposed OLDF algorithm surpasses the others in reducing the average voltage estimation error ($\varepsilon_{avg}$) across nearly all scenarios with the exception of the \texttt{22-bus} case. For this case, our OLDF results were better than all except the PLPF approximation where the average error was still quite close ($0.00020$~{p.u.} for PLPF vs. $0.00028$~{p.u.} for OLDF). Regarding the maximum errors in the high-load scenarios, no individual approximation consistently dominated the others across all test cases. However, we note that summing the maximum errors across all test cases reveals that the OLDF parameters lead to the best performance overall for this metric. These results show that OLDF parameters trained with scenarios around base-load conditions nevertheless perform well for high-load conditions. Furthermore, as an adaptive power flow, the OLDF approximation has the ability to tailor the parameters to perform even better for these conditions by including more training scenarios associated with high loading.

\subsubsection{Random Load Evaluation}
\label{subsec:Random Load Evaluation}
The OLDF approximation's accuracy is further analyzed for random loading conditions using $10,000$ scenarios sampled from a uniform distribution of 0\% to 150\% of the base load. 
Similar to the base and high load conditions, the performance metrics for the random load conditions detailed in Table~\ref{tab:comparison_Random} also show the OLDF approximation's dominance, with this approximation having the smallest max and average voltage magnitude errors ($\varepsilon_{\max}$ and $\varepsilon_{avg}$). 
Across all test cases, the OLDF accuracy improvement over traditional LDF ranges from $26.42\%$ to $91.67\%$ for average error (\(\varepsilon_{avg}\)) and from $20.94\%$ to $87.78\%$ for maximum error (\(\varepsilon_{max}\)). Compared to the best of PLPF and LoDF, OLDF's improvement ranges from $10\%$ to $80\%$ for average error and from $5.56\%$ to $80.70\%$ for max error.

\begin{table*}[t]
    \centering
    \caption{Model Evaluation - Random Load}
    \renewcommand{\arraystretch}{1.3}
    \label{tab:comparison_Random}
    \begin{tabular}{|p{0.055cm}>{\centering\arraybackslash}c|cccc|cccc|}
    \hline
&\text{Test case} & \(\varepsilon_{avg}^{LDF}\) & \(\varepsilon_{avg}^{PLPF}\) & \(\varepsilon_{avg}^{LoDF}\) & \(\varepsilon_{avg}^{OLDF}\) & \(\varepsilon_{\max}^{LDF}\) & \(\varepsilon_{\max}^{PLPF}\) & \(\varepsilon_{\max}^{LoDF}\) & \(\varepsilon_{\max}^{OLDF}\) \\
\hline \hline
\multirow{7}{*}{\rotatebox{90}{\textbf{\underline{$1\phi$}}}}
&22-bus & 0.00024 & 0.00010 & 0.00220 & $\mathbf{0.00002}$ & 0.00090 & 0.00057 & 0.00477 & $\mathbf{0.00011}$ \\
&33-bus & 0.00114 & 0.00051 & 0.00262 & $\mathbf{0.00019}$ & 0.00443 & 0.00312 & 0.00662 & $\mathbf{0.00124}$ \\
&69-bus & 0.00077 & 0.00051 & 0.00101 & $\mathbf{0.00017}$ & 0.00918 & 0.00816 & 0.00476 & $\mathbf{0.00347}$ \\
&85-bus & 0.00278 & 0.00056 & 0.00261 & $\mathbf{0.00031}$ & 0.00735 & 0.00321 & 0.00347 & $\mathbf{0.00257}$ \\
&123-bus & 0.00120 & 0.00080 & 0.00392 & $\mathbf{0.00072}$ & 0.00300 & 0.00227 & 0.00624 & $\mathbf{0.00139}$ \\
&141-bus & 0.00084 & 0.00035 & 0.00273 & $\mathbf{0.00013}$ & 0.00241 & 0.00108 & 0.00516 & $\mathbf{0.00102}$ \\
&533-bus & 0.00018 & - & - & \(\mathbf{0.00013}\) & 0.00097 & - & - & \(\mathbf{0.00061}\) \\
&906-bus &0.02548  &  - & - & \(\mathbf{0.00051}\) &0.02569  & - & - & \(\mathbf{0.00265}\) \\
\hdashline
\multirow{3}{*}{\rotatebox{90}{\textbf{\underline{$3\phi$}}}}
&13-bus & 0.02703 & - & - &  \(\mathbf{0.01335}\) &0.13930  & - & - & \(\mathbf{0.08872}\) \\
&37-bus &0.01438   & - & - & \(\mathbf{0.01058}\) &0.07639  & - & - & \(\mathbf{0.05453}\) \\
&123-bus &0.01274  & - & - & \(\mathbf{0.00637}\) &0.05455  & - & - & \(\mathbf{0.03506}\) \\

        \hline
    \end{tabular}
    \begin{tablenotes}
    \item[*] \scriptsize \hspace{10em}$1\phi$ stands for balanced single-phase equivalent networks, and $3\phi$ stands for unbalanced three-phase networks.
 \item[*] \scriptsize  \hspace{10em}The best performing method (smallest loss function) is bolded for each test case. All values are in per unit.
\end{tablenotes}
\label{table:Model evaluation- random load}
\end{table*}
\subsection{Computational Efficiency}
\label{subsec:Computational Efficiency}
As shown in Fig.~\ref{fig:flowchart}, 
our algorithm computes \mbox{LinDistFlow} parameters during an offline phase where ample computing time is available. These parameters are then used in online applications where computing time may be limited. Thus, the training process for our proposed algorithm requires computational tractability consistent with an offline context. As shown by the training times in Table~\ref{tab:comparison} that range from $0.6451$ seconds for the \texttt{22-bus} case to $25.3917$ seconds for the \texttt{123-bus} case, leveraging mature optimization methods like TNC enables acceptable scalability for the training phase. 

The computation times for online uses of the optimized \mbox{LinDistFlow} parameters depend on the particular application for which they are employed. However, since the only changes are to the parameter values and not the mathematical form of the \mbox{LinDistFlow} expressions, online computation times with our optimized parameters should be comparable to existing \mbox{LinDistFlow} approximations. 
To show this, the average calculation times for the $10,000$ random load scenarios in the experiment in Section~\ref{subsec:Random Load Evaluation} ranged from $0.0004$ to $0.1156$ seconds, with differences of less than $2$\% between the \mbox{LinDistFlow} using traditional parameters and our optimized parameters.

\begin{table}[t]
    \centering
    \footnotesize
    \caption{Computation Times in Seconds}
    \label{tab:comparison}
    \setlength{\tabcolsep}{1.5pt} 
    \renewcommand{\arraystretch}{1.5} 
    \scriptsize 
    \begin{tabularx}{\columnwidth}{X *{8}{>{\centering\arraybackslash}X} : *{3}{>{\centering\arraybackslash}X}}
    \toprule
    Test case & 22-  bus & 33-  bus & 69- 
 bus & 85-  bus & 123-bus & 141-bus & 533-bus & 906-bus & 13-  bus & 37-  bus& 123-bus\\ 
    \midrule
    $t_{train}$ & 0.6451 & 0.7852 & 4.9879 & 2.7182 & 2.3093 & 2.6942 & 3.4019 & 7.1058 & 1.2611 & 6.7772&25.3917\\
    \hdashline
    $t_{base}$ & 0.0071 & 0.0092 & 0.0161 & 0.0174 & 0.0112 & 0.0114 & 0.1025 & 0.3061 & 0.0006 & 0.0539&1.1495\\
    $t_{10000}$ & 0.0004 & 0.0005 & 0.0008 & 0.0009 & 0.0006 & 0.0014 & 0.0023 & 0.0042 & 0.0182 &0.0362&0.1156 \\
    \bottomrule
    \end{tabularx}
    \begin{tablenotes}
        \item[*] \scriptsize $t_{train}$: Offline computation time to train the parameters.
        \item[*] \scriptsize $t_{base}$: Online computation time for the base case loading (i.e., one scenario).
        \item[*] \scriptsize $t_{10000}$: Online per-scenario time averaged over $10,000$ random scenarios.
    \end{tablenotes}
\end{table}

\begin{figure}[!h]
    \centering
    \includegraphics[width=0.48\textwidth]{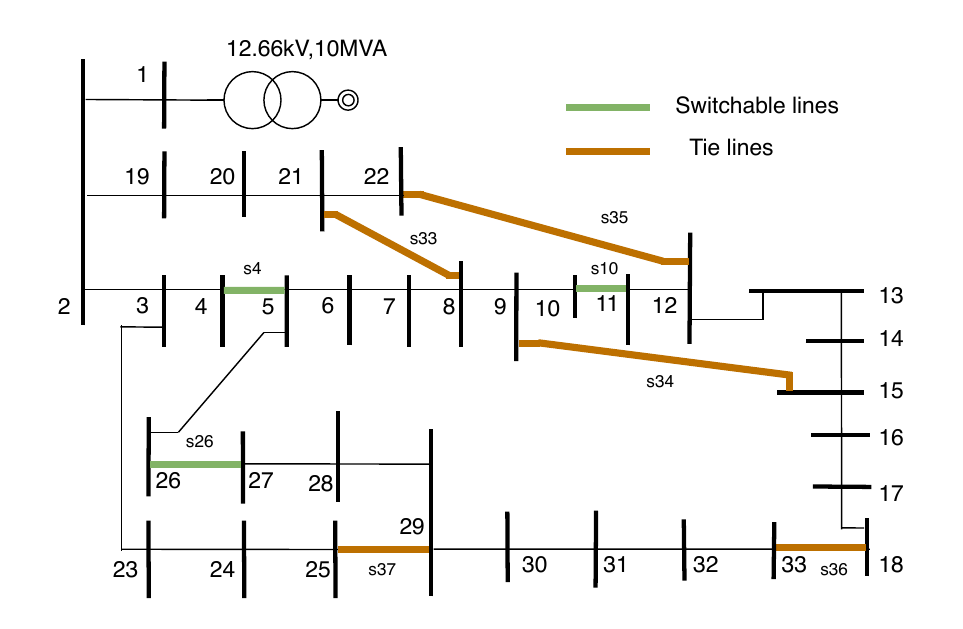}
    \caption{The \texttt{IEEE 33-bus} system with 3 switchable lines and 5 tie-lines.}
    \label{fig:33bus}
\end{figure}

\begin{table}[ht]
\centering
\caption{Topology Changes: Opened and Closed Lines}
\label{tab:topology_changes}
\begin{tabular}{ll|ll|ll}
\toprule
\textbf{Opened} & \textbf{Closed} & \textbf{Opened} & \textbf{Closed} & \textbf{Opened} & \textbf{Closed} \\
\midrule

- & - & (4, 10) & (34, 35) &(10, 26) & (35, 36) \\
(4) & (33) & (4, 10) & (34, 37) &(10, 26) & (35, 37) \\
(4) & (35) & (4, 10) & (35, 36) & (10, 26) & (36, 37) \\
(4) & (37)& (4, 10) & (35, 37) & (4, 10, 26) & (33, 34, 35)  \\
(10) & (34)& (4, 10) & (36, 37) & (4, 10, 26) & (33, 34, 36)  \\
(10) & (35)& (4, 26) & (33, 36) &  (4, 10, 26) & (33, 35, 36) \\
(10) & (36) & (4, 26) & (33, 37) &(4, 10, 26) & (33, 35, 37) \\
(26) & (36)& (4, 26) & (35, 36) & (4, 10, 26) & (33, 36, 37)  \\
(26) & (37)& (4, 26) & (35, 37) &  (4, 10, 26) & (34, 35, 36) \\
(4, 10) & (33, 34) & (4, 26) & (36, 37)  &(4, 10, 26) & (34, 35, 37) \\
(4, 10) & (33, 35)& (10, 26) & (34, 35)  &(4, 10, 26) & (34, 36, 37)  \\
(4, 10) & (33, 36) & (10, 26) & (34, 36) &\\

\bottomrule
\end{tabular}
   \begin{tablenotes}
 \item[*] \scriptsize These are all the valid switching combinations that lead to connected radial configurations for the \texttt{IEEE 33-bus} test case shown in Fig.~\ref{fig:33bus}. Note that the table indicates the changes from the configuration shown in Fig.~\ref{fig:33bus}. 
\end{tablenotes}
\end{table}

\begin{figure}
    \centering
    \includegraphics[width=0.49\textwidth]{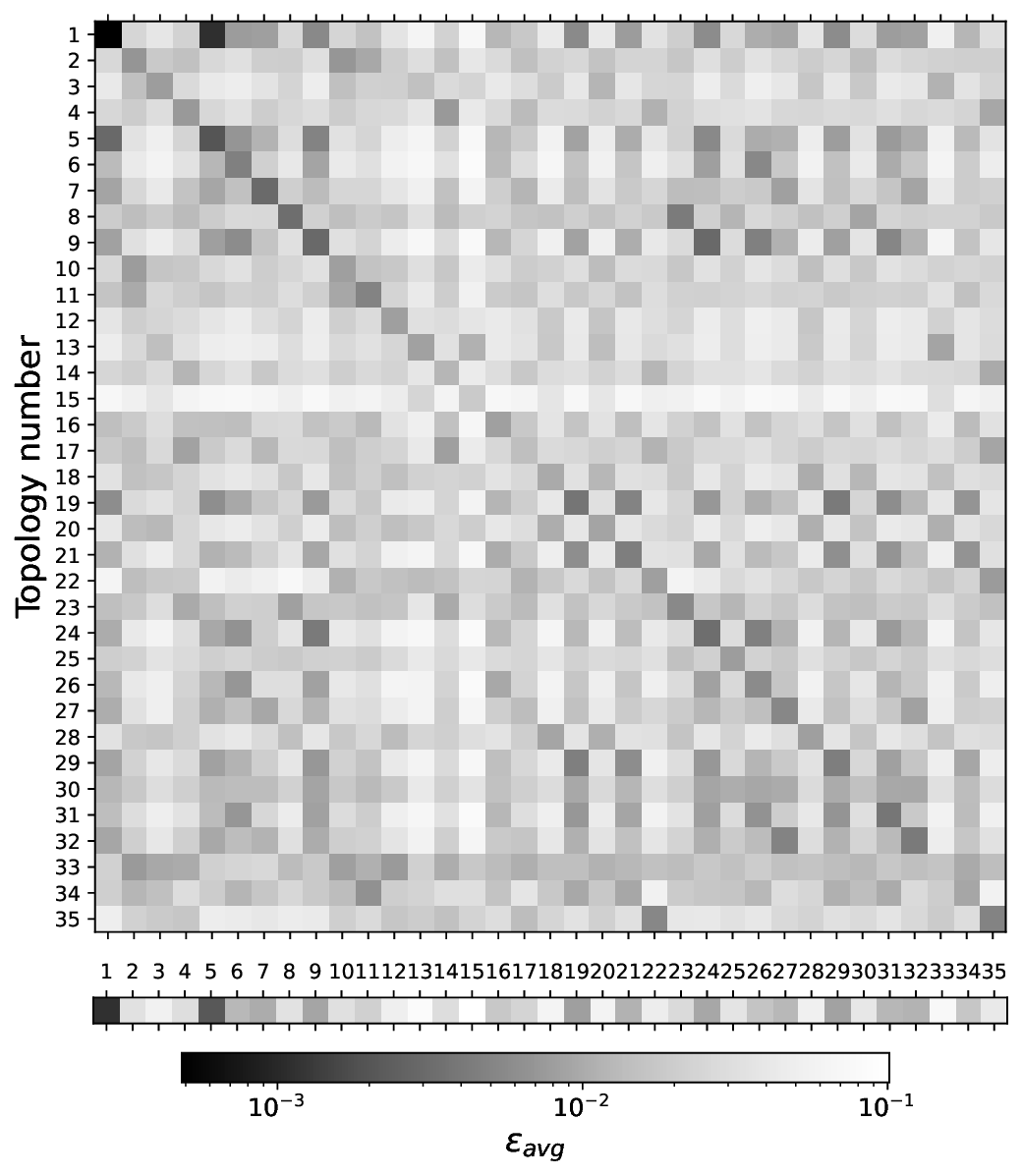}
    \caption{Heatmap visualization of average error metric ($\varepsilon_{avg}$) across 35 topologies for the \texttt{IEEE 33-bus} test case, with a log-scaled color representation to highlight performance variations. The accompanying vector plot below the heatmap quantitatively assesses the \mbox{LinDistFlow} approximation's baseline performance across the same topologies, facilitating a direct comparison of adaptability and optimization effectiveness.}
    \label{fig:heatmap}
\end{figure}

\subsection{Topology Analysis}

Engineers frequently need to both assess the impacts of topology changes for contingency assessments and optimize the topology of distribution systems for voltage management, loss minimization, outage restoration, etc.~\cite{mahdavi2021}. Thus, it is important to study the OLDF model's accuracy across topologies.

Addressing topology changes with the \mbox{LinDistFlow} approximation can be approached in various ways. For example, the traditional \mbox{LinDistFlow} approximation allows for straightforward topology adjustments by updating the $\mathbf{D}_{r}$ and $\mathbf{D}_{x}$ matrices with new values corresponding to the altered topology to calculate voltages. Similarly, in our optimization-based algorithm, adjustments can be made by excluding optimized parameters for removed lines and incorporating original resistance and reactance values for new lines in $\mathbf{D}_{r}^{opt}$ and $\mathbf{D}^{opt}_{x}$, without altering the bias parameters ($\boldsymbol{\gamma}^{opt}$, $\boldsymbol{\rho}^{opt}$, and $\boldsymbol{\varrho}^{opt}$). Successfully maintaining performance with this strategy indicates that our algorithm adapts well to different network topologies, avoiding overfitting to a specific configuration. Alternatively, optimizing parameters specifically for each topology through dedicated optimization could enhance accuracy at the cost of increased computational time and storage for the additional parameters needed. Nevertheless, such calculations could be efficiently executed in parallel for a specified set of topologies, making this process suitable for high-performance computing environments, as each topology's optimization process operates independently of others. 

To explore these different approaches, we describe a small-scale experiment. Fig.~\ref{fig:33bus} depicts the \texttt{IEEE 33-bus} distribution system, as described in~\cite{baran1989network}, which has $33$ nodes and $37$ lines. 
We consider a version of this system with eight switchable lines, specifically, lines $4$, $10$, and $26$ are normally closed switches (NCS) and lines $33$ to $37$ are tie lines or normally open switches (NOS).
With these switchable lines, we can create $35$ distinct and valid (i.e., radial and connected) topologies out of $56$ possible topologies, as listed in Table~\ref{tab:topology_changes}.

\begin{figure}
    \centering
    \includegraphics[width=0.49\textwidth]{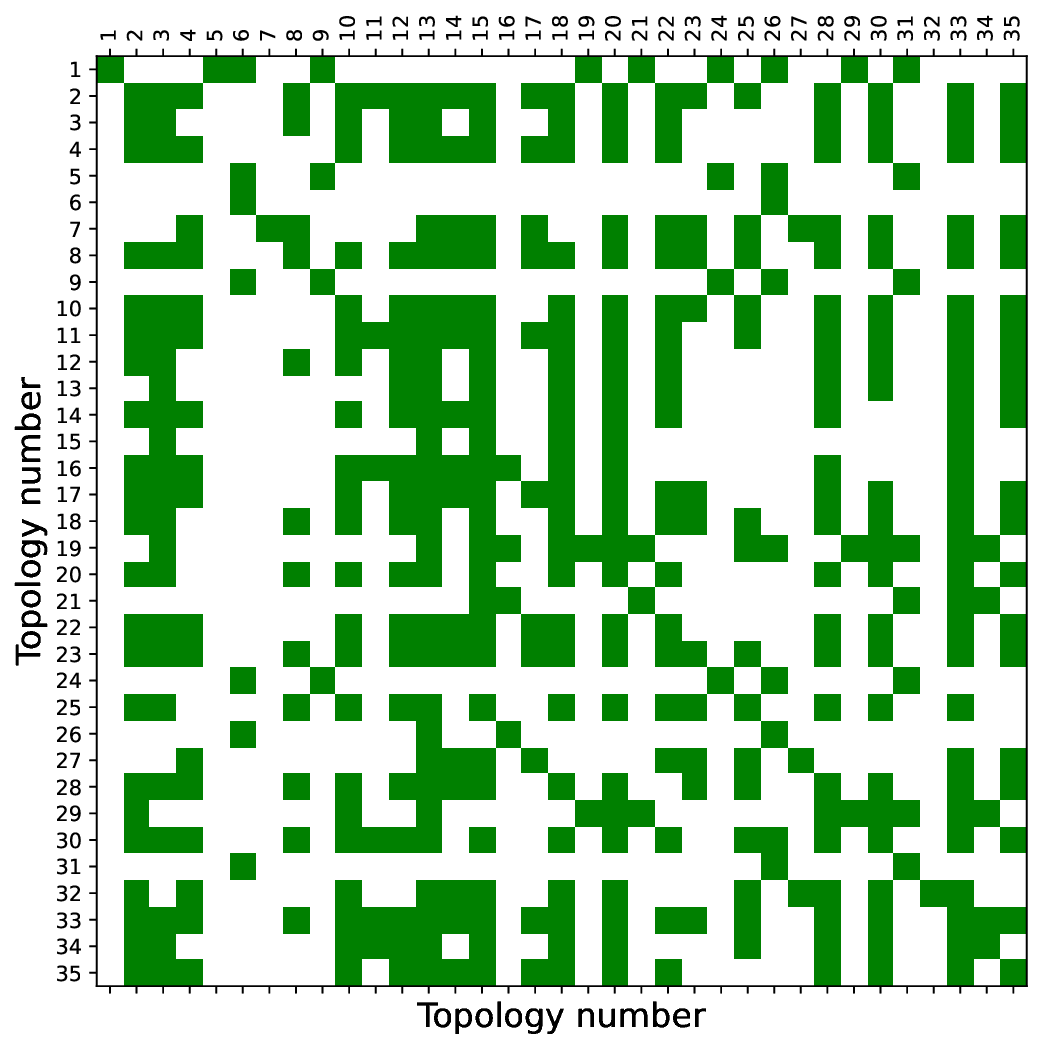}
    \caption{Binary comparison matrix between the traditional \mbox{LinDistFlow} approximation's performance and optimized parameters from our proposed algorithm across 35 topologies in the \texttt{IEEE 33-bus} test case. Each column represents a topology, with green indicating that the optimized parameters for the topology associated with the corresponding row outperform the traditional \mbox{LinDistFlow} approximation and white indicating that the traditional \mbox{LinDistFlow} approximation performance is better.}
    \label{fig:heatmap_comp}
\end{figure}

We next evaluate the adaptability of optimized parameters, i.e., assess how well parameters optimized for one topology perform on others. To accomplish this, we performed the offline training phase of our proposed algorithm using the same setup as before across the $35$ network topologies to obtain a dataset with $35$ sets of optimized coefficients and bias parameters. We tested the performance of these optimized parameters using the same $10,000$ test samples as before on each topology.

Illustrating this cross-topology assessment, Fig.~\ref{fig:heatmap} shows the performance of optimized parameters considering all $35$ topologies. Specifically, the heatmap at the top of this figure shows the average error metric, $\varepsilon_{avg}$,  of the optimized parameters for a given topology (rows) when applied to different topologies (columns). The horizontal vector plot at the bottom of the figure shows the performance of the traditional \mbox{LinDistFlow} approximation on the same $35$ topologies. The heatmap employs a logarithmic scale for color representation to enable comparisons across a broad range of error magnitudes. The results underscore the variability in the algorithm's adaptability, with darker shades indicating lower errors (better performance) and lighter shades denoting higher errors. The vector plot beneath the heatmap contrasts the overall baseline performance of the traditional \mbox{LinDistFlow} approximation.

The matrix in Fig.~\ref{fig:heatmap_comp} further illustrates the performance of our optimized parameters relative to the traditional \mbox{LinDistFlow} approximation (i.e., comparing each row from heatmap in Fig.~\ref{fig:heatmap} to the horizontal vector plot at the bottom of the figure). This matrix employs a binary color coding---green for topologies where the optimized parameters outperform the traditional \mbox{LinDistFlow} approximation and white where they do not. For example, row $1$ in this matrix shows that training the parameters using the base topology results in the optimized parameters outperforming the traditional \mbox{LinDistFlow} approximation on topologies $1$, $5$, $6$, $9$, $19$, $21$, $24$, $26$, $29$, and $31$. As another example, training the parameters using topology $2$ leads  to superior performance on topologies $\{[2,4] \cup 8 \cup [10,15] \cup [17,18] \cup 20 \cup [22,23] \cup 25 \cup 28 \cup 30 \cup 33 \cup 35\}$. 

These visualizations show the nuanced performance of the optimized parameters vis-\`a-vis the traditional \mbox{LinDistFlow} approximation across topologies. Our future work aims to cluster topologies for which jointly optimized parameter values can provide accurate \mbox{LinDistFlow} approximations.

\textcolor{black}{The adaptability of the OLDF framework also extends to modeling active control devices like switched capacitors and voltage regulators. In scenarios where these devices are in a fixed configuration or are locally controlled, the offline training process has an advantage over traditional methods by implicitly learning the average impact of their behavior from the power flow solution samples. For optimization problems where device settings are decision variables, the OLDF model is also well-suited. The reactive power injections from capacitor banks can be treated as direct inputs ($q_n$) to the model. Furthermore, different tap positions of a voltage regulator can be modeled as distinct network configurations, allowing for tailored OLDF parameters to be trained offline for each setting, leveraging the same parallelizable approach used for topology changes. This flexibility allows the OLDF model to create more accurate, system-specific linearizations for a wider range of practical operating scenarios.}

\subsection{Illustrative Application: Hosting Capacity Computation}
As an illustrative example application, this section demonstrates the use of the proposed optimized \mbox{LinDistFlow} model to determine the hosting capacity of inverter-based generation units. Following~\cite{zhan2023fairness}, the hosting capacity problem is:

\begin{subequations}
\label{eq:HC}
    \begin{align}
     & \minimize_{p_{n}^{g}, q_{n}^{g}} \quad  \sum_{n \in \mathcal{\hat{N}}} \left( \frac{(\overline{p_{n}^{g}}- p_{n}^{g})^{2}} {\overline{p_{n}^{g}}} + \xi \frac{{q_{n}^{g}}^{2}} {\overline{s_{n}^{g}}} \right) \label{eq:HC_objective} \\
    &  \text{s.t.} \quad  0 \leq p_{n}^{g} \leq \overline{p_{n}^{g}}, \quad \quad |q_{n}^{g}| \leq \sqrt{\overline{{s_{n}^{g}}^2} - {p_{n}^{g}}^2}, ~\forall n \in \mathcal{\hat{N}}\label{eq:HC_p_q_cons} \\
    &\quad \quad v_{0}= 1, \quad \underline{v_{n}} \leq v_{n} \leq \overline{v_{n}}, \hspace{1.8cm} \forall n \in \mathcal{N}^{\prime} \label{eq:HC_v_cons} \\
    &\quad  \quad \sqrt{P^{2}_{n}+Q^{2}_{n}} \leq S_{n}, \hspace{1.45cm}\forall n:(\pi_n, n) \in \mathcal{E} \label{eq:HC_line_cons} \\
    &\quad \quad \sqrt{P_{T}^{2}+Q_{T}^{2}} \leq S_{T}, \label{eq:HC_transformer1} \\
    &\quad \quad P_{T} = \sum_{n:(0,n)\in \mathcal{E}}  P_{0n}  ,  \quad Q_{T} = \sum_{n:(0,n)\in \mathcal{E}}  Q_{0n},\label{eq:HC_transformer2}\\
    & \quad \quad \text{LDF \eqref{eq:matrixform} or OLDF \eqref{eq:voltage_final}},
\label{eq:HC_voltage}
    \end{align}
\end{subequations}
where $\mathcal{\hat{N}}$ is the set of buses with inverter-based generators, $p_{n}^{g}$ and $q_{n}^{g}$ are the active and reactive power generation, $\overline{p_{n}^{g}}$ and $\overline{s_{n}^{g}}$ are their maximum capacities, $S_n$ and $S_T$ are the upper bounds on the line and transformer flows, and $\xi$ is a weighting factor controlling the tradeoff between active power and reactive power utilization. The constraints ensure adherence to power generation limits~\eqref{eq:HC_p_q_cons} and voltage regulation requirements~\eqref{eq:HC_v_cons} along with line~\eqref{eq:HC_line_cons} and transformer capacities~\eqref{eq:HC_transformer1}--\eqref{eq:HC_transformer2}.

An evaluation using the \texttt{IEEE 33-bus} test system illustrates the OLDF model's effectiveness. We set $\overline{s_{n}^{g}} = 0.6$ MVA with a $0.98$ power factor, $\xi=0.02$, and voltage limits between $1.05$ and $0.95$ per unit, with a substation capacity of $10$ MVA. 

\begin{figure}[ht]
    \centering
    \includegraphics[width=0.49\textwidth]{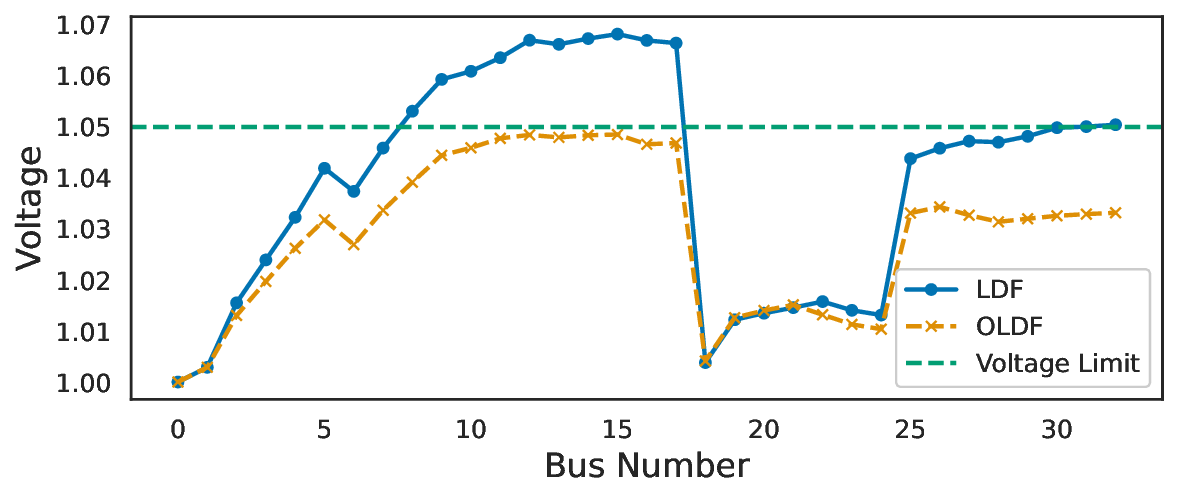}
    \caption{Voltage profile of the \texttt{IEEE 33-bus} test case after feeding the active/reactive power injections obtained from solving \eqref{eq:HC} with LDF (blue curve) and OLDF (orange curve) into the \mbox{DistFlow} equations.}
    \label{fig:hosting}
    \vspace{-1em}
\end{figure}

Upon solving \eqref{eq:HC} with both LDF and \textcolor{black}{our proposed} OLDF models, we obtain the optimal active and reactive power settings for the inverter-based generation units. By assessing these optimal settings using the original nonlinear \mbox{DistFlow} model, we compare the performance of the OLDF and LDF approximations. \textcolor{black}{To clarify the comparison methodology, we first solved the optimization problem \eqref{eq:HC} using the LDF model to find its proposed optimal generator setpoints. We then solved the same problem using the OLDF model to find a different set of optimal setpoints. Finally, to determine the true feasibility and performance of each solution, we evaluated \textit{both} sets of setpoints using the accurate nonlinear DistFlow model. Fig.~\ref{fig:hosting} plots the actual voltage profiles resulting from this validation.} As shown in this figure, while the traditional \mbox{LinDistFlow} approximation leads to voltage violations at certain buses within the hosting capacity problem \eqref{eq:HC}, the application of the proposed OLDF model avoids such violations. 
\textcolor{black}{The voltage violations resulting from the LDF-based optimization in Fig.~\ref{fig:hosting} occur because traditional LDF inaccurately models the system's voltage response to the high generation injections inherent in hosting capacity analysis. This can lead LDF to identify generation setpoints that appear feasible within its simplified model but are, in fact, infeasible in the true nonlinear system. Conversely, OLDF's parameters, having been optimized for broad accuracy across diverse conditions, provide a more accurate system representation, thus yielding genuinely feasible solutions. Such discrepancies, where LDF-based decisions prove unreliable when operating near system limits, are generally anticipated given OLDF's consistently superior performance demonstrated across various test cases and loading scenarios (Tables~\ref{table:Model evaluation- base load}, \ref{tab:comparison_High}, and~\ref{tab:comparison_Random}).
}

%% file: text/appendix.tex
\appendix[Derivation of Three-Phase LinDistFlow (LinDist3Flow) Model]\label{appendix:three-phase_derivations}

This appendix presents a detailed derivation of the linearized three-phase power flow model, LinDist3Flow~\cite{gan2014convex, arnold2016optimal}. Specifically, this appendix derives the voltage update equation and the coefficient matrices $\mathbb{H}^P$ and $\mathbb{H}^Q$ that are presented in equations~\eqref{eq:voltage_final_3phase},~\eqref{eq:22}, and~\eqref{eq:23} in Section~\ref{sec:Multi-Phase Grids} of the main paper.
Consider two adjacent buses in the distribution feeder, denoted as \( (n,k) \in \mathcal{N} \). We first express the power flow equations in vector form, as follows:
\begin{align}
    \mathtt{v}_n &= \mathtt{v}_k + \mathtt{Z}_k  \mathtt{I}_k, \label{eq:voltage_eq}\\
    \mathtt{I}_n &= \mathtt{i}_n + \sum_{k: n \rightarrow k} \mathtt{I}_k, \label{eq:current_eq}
\end{align}
where $\mathtt{v}_n = \begin{bmatrix} V^a & V^b & V^c\end{bmatrix}_{n}^{\top}$ is the vector of voltage phasors at node $n$, $\mathtt{i}_n = \begin{bmatrix} i^a & i^b & i^c\end{bmatrix}_{n}^{\top}$ is the vector of load currents at node $n$, $\mathtt{I}_n = \begin{bmatrix} I^a & I^b & I^c\end{bmatrix}_{n}^{\top}$ is the vector of current phasors entering node $n$. Next, we multiply both sides of \eqref{eq:voltage_eq} by their complex conjugates and similarly multiply both sides of \eqref{eq:current_eq} by \( \mathtt{v}_n \). This gives us the following equations:
\begin{align}
    \mathtt{v}_n \mathtt{v}_n^* &= \mathtt{v}_k \mathtt{v}_k^* + \mathtt{Z}_k \mathtt{I}_k \mathtt{v}_k^* + \mathtt{v}_k \mathtt{I}_k^* \mathtt{Z}_k^* + \mathtt{Z}_k \mathtt{I}_k \mathtt{I}_k^* \mathtt{Z}_k^*, \nonumber\\
              &= \mathtt{v}_k \mathtt{v}_k^* + 2\Re \{ \mathtt{v}_k \mathtt{I}_k^* \mathtt{Z}_k^* \} + \mathtt{Z}_k \mathtt{I}_k \mathtt{I}_k^* \mathtt{Z}_k^*, \label{eq:voltage_squared}\\
    \mathtt{v}_n \mathtt{I}_n^* &= \mathtt{v}_n \mathtt{i}_n^* + \sum_{k: n \rightarrow k} (\mathtt{v}_k + \mathtt{Z}_k \mathtt{I}_k) \mathtt{I}_k^*. \label{eq:current_power_eq}
\end{align}

Similar to the derivation of the LinDistFlow model, we can ignore the loss terms in \eqref{eq:voltage_squared} and \eqref{eq:current_power_eq}, which results in:
\begin{align}
    \mathtt{v}_n \mathtt{v}_n^* &\approx \mathtt{v}_k \mathtt{v}_k^* + 2 \Re \{ \mathtt{v}_k \mathtt{I}_k^* \mathtt{Z}_k \}, \label{eq:linear_voltage_squared}\\
    \mathtt{v}_n \mathtt{I}_n^* &\approx \mathtt{v}_n \mathtt{i}_n^* + \sum_{k: n \rightarrow k} \mathtt{v}_k \mathtt{I}_k^*. \label{eq:linear_current_power}
\end{align}
Equations \eqref{eq:linear_voltage_squared} and \eqref{eq:linear_current_power} represent $3 \times 3$ matrix equations. To focus on \eqref{eq:linear_current_power}, we apply the power equation \( S_{\phi,k} = V_{\phi,k} I_{\phi,k}^* \) and collect the diagonal terms into a vector equation where \( S_{\phi,k}\) is the complex power phasor entering node $k$ at phase $\phi$:
\begin{equation}
    \mathtt{S}_n \approx \mathtt{s}_n + \sum_{k: n \rightarrow k} \mathtt{S}_k,\label{eq:power_sum}
\end{equation}
where $\mathtt{s}_n$ is the vector of complex loads at node $n$. Now, returning to \eqref{eq:linear_voltage_squared}, we can expand \( \mathtt{I}_k \) using the power equation \( S_{\phi,k} = V_{\phi,k} I_{\phi,k}^* \), which yields:
\begin{align}
    \mathtt{v}_n \mathtt{v}_n^* &\approx \mathtt{v}_k \mathtt{v}_k^* + 2 \Re \left\{ \mathtt{v}_k \left[ S_a v_a^{-1} S_b v_b^{-1} S_c v_c^{-1} \right]_{nk} \mathtt{Z}_{k}^* \right\}, \label{eq:expanded_voltage_squared}
\end{align}
\begin{align}
    \mathtt{v}_n \mathtt{v}_n^* &\approx \mathtt{v}_k \mathtt{v}_k^* \nonumber\\ + &2 \Re \left\{ 
    \begin{bmatrix}
        S_a & V_a S_b V_b^{-1} & V_a S_c V_c^{-1}\\
        V_b S_a V_a^{-1}& S_b & V_b S_c V_c^{-1}\\
        V_c S_a V_a^{-1}& V_ac S_b V_b^{-1} & S_c
    \end{bmatrix} \mathtt{Z}_{nk}^* \right\}_{k}. \label{eq:simplified_voltage_squared}
\end{align}

We now assume the ratios of voltage phasors are constant:
\begin{align}
    V_{a,k} V_{b,k}^{-1} &\approx \alpha, \quad V_{b,k} V_{c,k}^{-1} \approx \alpha, \quad V_{a,k} V_{c,k}^{-1} \approx \alpha^{2}. \label{eq:voltage_phasor_ratios}
\end{align}
Here, the constants \( \alpha \) and \( \alpha^2 \) are defined as:
\begin{align}
    \alpha &= 1 \angle 120^\circ = -\frac{1}{2} + \mathrm{j} \frac{\sqrt{3}}{2}, \quad \alpha^2 = 1 \angle 240^\circ = -\frac{1}{2} - \mathrm{j} \frac{\sqrt{3}}{2}. \label{eq:alpha_values}
\end{align}

Substituting these values from \eqref{eq:voltage_phasor_ratios} and \eqref{eq:alpha_values} into \eqref{eq:simplified_voltage_squared}, we obtain the final form:
\begin{align}
    \mathtt{v}_n \mathtt{v}_n^* &\approx \mathtt{v}_k \mathtt{v}_k^* + 2 \Re \left\{ 
    \begin{bmatrix}
        S_a & \alpha S_b & \alpha^2 S_c \\
        \alpha^2 S_a & S_b & \alpha S_c \\
        \alpha S_a & \alpha^2 S_b & S_c
    \end{bmatrix}_{k} \mathtt{Z}_{nk}^* \right\}. \label{eq:simplified_voltage_squared2}
\end{align}

We are mainly concerned with the diagonal elements of~\eqref{eq:simplified_voltage_squared2}, which we gather into a $3 \times 1$ vector:
\begin{align}
    \mathbb{V}_n &\approx \mathbb{V}_k \nonumber \\&+ 2 \Re \left\{ 
    \begin{bmatrix}
        Z_{aa,nk}^* S_{a,k} + \alpha Z_{ab,nk}^* S_{b,k} + \alpha^2 Z_{ac,nk}^* S_{c,k} \\
        \alpha^2 Z_{ba,nk}^* S_{a,k} + Z_{bb,nk}^* S_{b,k} + \alpha Z_{bc,nk}^* S_{c,k} \\
        \alpha Z_{ca,nk}^* S_{a,k} + \alpha^2 Z_{cb,nk}^* S_{b,k} + Z_{cc,nk}^* S_{c,k}
    \end{bmatrix} \right\}, \label{eq:diagonal_voltage}
\end{align}
\begin{align}
    \mathbb{V}_n &\approx \mathbb{V}_k + 2 \Re \left\{ 
    \begin{bmatrix}
        Z_{aa}^* + \alpha Z_{ab}^* + \alpha^2 Z_{ac}^*  \\
        \alpha^2 Z_{ba}^*  + Z_{bb}^*+ \alpha Z_{bc}^* \\
        \alpha Z_{ca}^* + \alpha^2 Z_{cb}^* + Z_{cc}^*
    \end{bmatrix}_{nk} \begin{bmatrix}
        S_a \\ S_b \\S_c
        \end{bmatrix}_{k}\right\}, 
    \label{eq:diagonal_voltage2}
\end{align}
where $\mathbb{V}_n = \begin{bmatrix} v^a & v^b & v^c\end{bmatrix}_{n}^{\top}$ is the vector of squared of voltage magnitudes for each phase $(a, b, c)$ at node $n$. Expanding the impedance matrix entries \( Z_{\phi\psi,k} = r_{\phi\psi,k} + \mathrm{j}x_{\phi\psi,k} \), and using \( S_{\phi,k} = P_{\phi,k} + \mathrm{j}Q_{\phi,k} \), we simplify \eqref{eq:diagonal_voltage} to:
\begin{equation}
    \mathbb{V}_n \approx \mathbb{V}_k - \mathbb{H}_{nk}^P \mathbb{P}_k - \mathbb{H}_{nk}^Q \mathbb{Q}_k, \label{eq:linearized_voltage}
\end{equation}

where the matrices \( \mathbb{H}_{nk}^P \) and \( \mathbb{H}_{nk}^Q \) are defined as:
\begin{align}
    \mathbb{H}_{nk}^P &= \begin{bmatrix}
        -2r_{aa} & r_{ab} - \sqrt{3}x_{ab} & r_{ac} + \sqrt{3}x_{ac} \\
        r_{ba} + \sqrt{3}x_{ba} & -2r_{bb} & r_{bc} - \sqrt{3}x_{bc} \\
        r_{ca} - \sqrt{3}x_{ca} & r_{cb} + \sqrt{3}x_{cb} & -2r_{cc}
    \end{bmatrix}, \label{eq:M_P}\\
    \mathbb{H}_{nk}^Q &= \begin{bmatrix}
        -2x_{aa} & x_{ab} + \sqrt{3}r_{ab} & x_{ac} - \sqrt{3}r_{ac} \\
        x_{ba} - \sqrt{3}r_{ba} & -2x_{bb} & x_{bc} + \sqrt{3}r_{bc} \\
        x_{ca} + \sqrt{3}r_{ca} & x_{cb} - \sqrt{3}r_{cb} & -2x_{cc}
    \end{bmatrix}. \label{eq:M_Q}
\end{align}

Finally, we restate \eqref{eq:power_sum} for clarity:
\begin{equation}
    S_n \approx s_n + \sum_{k: n \rightarrow k} S_k. \label{eq:final_power_sum}
\end{equation}

Equations~\eqref{eq:linearized_voltage}--\eqref{eq:final_power_sum} represent a linearized model for unbalanced three-phase power flow, which maps the active and reactive power injections at node \( k \) into squared voltage magnitude differences. Contributions from all phases influence each other's squared voltage magnitudes, and reducing this system to a single-phase network recovers the classical LinDistFlow model, as noted in \cite{baran1989optimal2}.

Now we can write the \eqref{eq:linearized_voltage} in matrix form as follows:
\begin{align}
    \label{eq:voltage_final_3phase_1}
    \mathbb{V} &=  \mathbb{V}_{0} \mathbf{1} + \mathbb{A}_{3}^{-1}\text{bdiag}(\mathbb{H}^{P}) \mathbb{A}_{3}^{-\top} \mathbb{P} +
    \mathbb{A}_{3}^{-1}\text{bdiag}(\mathbb{H}^{Q}) \mathbb{A}_{3}^{-\top} \mathbb{Q},
\end{align}
where $\mathbb{V} = \begin{bmatrix} v^a_1 & v^b_1 & v^c_1 & \cdots & v^a_n & v^b_n & v^c_n \end{bmatrix}^{\top}$ is the vector of squared of voltage magnitudes for each phase $(a, b, c)$, $\mathbb{A}_{3}$ is the network incidence matrix, $\text{bdiag}(\,\cdot\,)$ is the block diagonal operator, and $\mathbb{P} = \begin{bmatrix} p^a_1 & p^b_1 & p^c_1 & \cdots & p^a_n & p^b_n & p^c_n \end{bmatrix}^{\top}$ and $\mathbb{Q} = \begin{bmatrix} q^a_1 & q^b_1 & q^c_1 & \cdots & q^a_n & q^b_n & q^c_n \end{bmatrix}^{\top}$ are the active and reactive power demand vectors.

Adding parameters $\boldsymbol{\rho}_{3}$, $\boldsymbol{\varrho}_{3}$, and $\boldsymbol{\gamma}_{3}$ to~\eqref{eq:voltage_final_3phase_1} yields:
\begin{align}
    \label{eq:voltage_final_3phase_2}
    \mathbb{V} &=  \mathbb{V}_{0} \mathbf{1} + \mathbb{A}_{3}^{-1}\text{bdiag}(\mathbb{H}^{P}) \mathbb{A}_{3}^{-\top} (\mathbb{P} +  \boldsymbol{\rho}_{3}) + \nonumber\\ 
    & \qquad \qquad + \mathbb{A}_{3}^{-1}\text{bdiag}(\mathbb{H}^{Q}) \mathbb{A}_{3}^{-\top} (\mathbb{Q} + \boldsymbol{\varrho}_{3}) + \boldsymbol{\gamma}_{3}.
\end{align}